\documentclass{sig-alternate-2013}

\usepackage{balance}  
\usepackage{graphics} 
\usepackage{graphicx}
\usepackage{caption}
\usepackage{times}    
\usepackage{url}      

\usepackage{subfigure}
\usepackage{epstopdf}
\usepackage{epsfig}
\usepackage{amsmath}
\usepackage{url}
\usepackage{enumitem}

\newcommand{\remove}[1]{}

\newcommand{\specialcell}[2][c]{
  \begin{tabular}[#1]{@{}c@{}}#2\end{tabular}}

\makeatletter
\def\url@leostyle{%
  \@ifundefined{selectfont}{\def\UrlFont{\sf}}{\def\UrlFont{\small\bf\ttfamily}}}
\makeatother
\def\pprw{8.5in}
\def\pprh{11in}

\usepackage[pdftex]{hyperref}
\hypersetup{
pdftitle={},
pdfauthor={LaTeX},
bookmarksnumbered,
pdfstartview={FitH},
colorlinks,
citecolor=black,
filecolor=black,
linkcolor=black,
urlcolor=black,
breaklinks=true,
}

\begin{document}

\title{Portrait of an Online Shopper: \\Understanding and Predicting Consumer Behavior}
\numberofauthors{3}
\author{
\alignauthor
Farshad Kooti, \\ Kristina Lerman\\
       \large USC Information Sciences Institute \\
       \large Marina del Rey, CA \\
       \large \{kooti, lerman\}@isi.edu
\alignauthor
Luca Maria Aiello\\
       \large Yahoo Labs \\
       \large London, UK \\
         \large alucca@yahoo-inc.com
\alignauthor
Mihajlo Grbovic, \\Nemanja Djuric,\\ Vladan Radosavljevic\\
        \large Yahoo Labs \\
        \large Sunnyvale, CA \\
        \large \{mihajlo, djurikom, vladan\}@yahoo-inc.com
}
\date{\today}

\maketitle

\begin{abstract}
Consumer spending accounts for a large fraction of the US economic activity. Increasingly, consumer activity is moving to the web, where digital traces of shopping and purchases provide valuable data about consumer behavior. We analyze these data extracted from emails and combine them with demographic information to characterize, model, and predict consumer behavior. Breaking down purchasing by age and gender, we find that the amount of money spent on online purchases grows sharply with age, peaking in late 30s. Men are more frequent online purchasers and spend more money when compared to women. Linking online shopping to income, we find that shoppers from more affluent areas purchase more expensive items and buy them more frequently, resulting in significantly more money spent on online purchases. We also look at dynamics of purchasing behavior and observe daily and weekly cycles in purchasing behavior, similarly to other online activities.

More specifically, we observe temporal patterns in purchasing behavior suggesting shoppers have finite budgets: the more expensive an item, the longer the shopper waits since the last purchase to buy it. We also observe that shoppers who email each other purchase more similar items than socially unconnected shoppers, and this effect is particularly evident among women. Finally, we build a model to predict when shoppers will make a purchase and how much will spend on it. We find that temporal features improve prediction accuracy over competitive baselines. A better understanding of consumer behavior can help improve marketing efforts and make online shopping more pleasant and efficient.


\end{abstract}

\section{Introduction}\label{section:intro}


\noindent
Consumer spending is an integral component of economic activity. In 2013, it accounted for 71\% of the US gross domestic product (GDP)\footnote{\url{https://research.stlouisfed.org/fred2/series/PCE/}}, a measure often used to quantify economic output and general prosperity of a country. Given its importance, many studies focused on understanding and characterizing consumer behavior. Researchers examined gender differences and motivations in shopping~\cite{goingshopping,JOCA:JOCA113}, as well as spending patterns across urban areas~\cite{sobolevsky15cities}.

In recent years, shopping has increasingly moved online. Consumers use internet to research product features, compare prices, and then purchase products from online merchants, such as Amazon and Walmart. Moreover, platforms like eBay allow people to directly sell products to one another. While there exist concerns about the risks and security of online shopping~\cite{bhatnagar00risk,perea04drives,teo02attitudes}, large numbers of people, especially younger and wealthier~\cite{horrigan2008online,swinyard03people}, choose online shopping even when similar products can be purchased offline~\cite{farag07shopping}.
In fact, online shopping has grown significantly, with an estimated \$1,471 billion dollars spent online in 2014 in the United States alone by 191 million online shoppers\footnote{\url{http://www.statista.com/topics/871/online-shopping/}}.


Most of these online purchases result in a confirmation or shipment email sent to the shopper by the merchant. These emails provide a rich source of evidence to study online consumer behavior across different shopping websites. Unlike previous studies~\cite{pavlou06understanding}, which were based on surveys and thus limited to relatively small populations, we used email data to perform a large-scale study of online shopping. Specifically, we extracted information about 121 million purchases amounting to 5.1B dollars made by 20.1 million shoppers, who are also Yahoo Mail users. The information we extracted included names of purchased products, their prices, and purchase timestamps. We used email user profile to link this information to demographic data, such as gender, age, and zip code. This information enabled us to characterize patterns of online shopping activity and their dependence on demographic and socio-economic factors. We found that, for example, men generally make more purchases and spend more on online purchases. Moreover, online shopping appears to be widely adopted by all ages and economic classes, although shoppers from more affluent areas generally buy more expensive products than less affluent shoppers.

Looking at temporal factors affecting online shopping, we found patterns similar to other online activity~\cite{Kooti2015www}. Not surprisingly, online shopping has daily and weekly cycles showing that people fit online shopping routines into their everyday life. However, purchasing decisions appear to be correlated. The more expensive a purchase, the longer the shopper has to wait since the previous purchase to buy it. This can be explained by the fact that most shoppers have a finite budget and have to wait longer between purchases to buy more expensive items.

In addition to temporal and demographic factors, social networks are believed to play an important role in shaping consumer behavior, for instance by spreading information about products through the word-of-mouth effect~\cite{rodrigues11word}. Previous studies examined how consumers use their online social networks to gather product information and recommendations~\cite{guo11role,gupta14identifying}, although the direct effect of recommendations on purchases was found to be weak~\cite{leskovec07dynamics}. In addition, people who are socially connected are generally more similar to one another than unconnected people~\cite{mcpherson01bird}, and hence, they are more likely to be interested in similar products. Our analysis confirmed that shoppers who are socially connected (because they email each other) tend to purchase more similar products than unconnected shoppers.

Once we understand the factors affecting consumer behavior, we are then able to predict it. Given users' purchase history and demographic data, we attempt to predict the time of their next purchase, as well as the price of their next purchase. Our method attains a relative improvement of at least 49.8\% over the random baseline for predicting the price of the next purchase and 36.4\% relative improvement over the random baseline for predicting the time of the next purchase. Interestingly, demographic features were the least useful in the prediction task, while temporal features carried the most discriminative information. 

The contributions of the paper are summarized below:
\begin{itemize}[leftmargin=*]
\itemsep0em
\item Introduction of a unique and very rich data set about consumer behavior, extracted from purchase confirmations merchants send to buyers (Section~\ref{section:dataset});
\item A quantitative analysis of the impact of demographic, temporal, and network factors on consumer behavior (Section~\ref{section:behavior});
\item Prediction of consumer behavior, specifically predicting when will the next purchase occur and how much money will be spent (Section~\ref{section:prediction}).
\end{itemize}

A better understanding of consumer behavior can benefit consumers, merchants, as well as advertisers. Knowing when consumers are ready to spend money and how much they are willing to spend can improve the effectiveness of advertising campaigns, and prevent consumers from wasting their resources on unnecessary purchases. Understanding these patterns can help make online shopping experience more efficient for consumers. Considering that consumer spending is such a large portion of the economy, even a small efficiency gain can have dramatic consequences on the overall economic activity.

\section{Dataset}\label{section:dataset}

Most online purchases result in a confirmation email being sent by the seller to the shopper. These emails provide an unique opportunity to study the shopping behavior of people across different online retail stores, such as Amazon, eBay, and Walmart.

\begin{figure}[t!]
\begin{tabular}{@{}c@{}c@{}}
\subfigure[PDF]{
   \includegraphics[width=0.5\columnwidth]{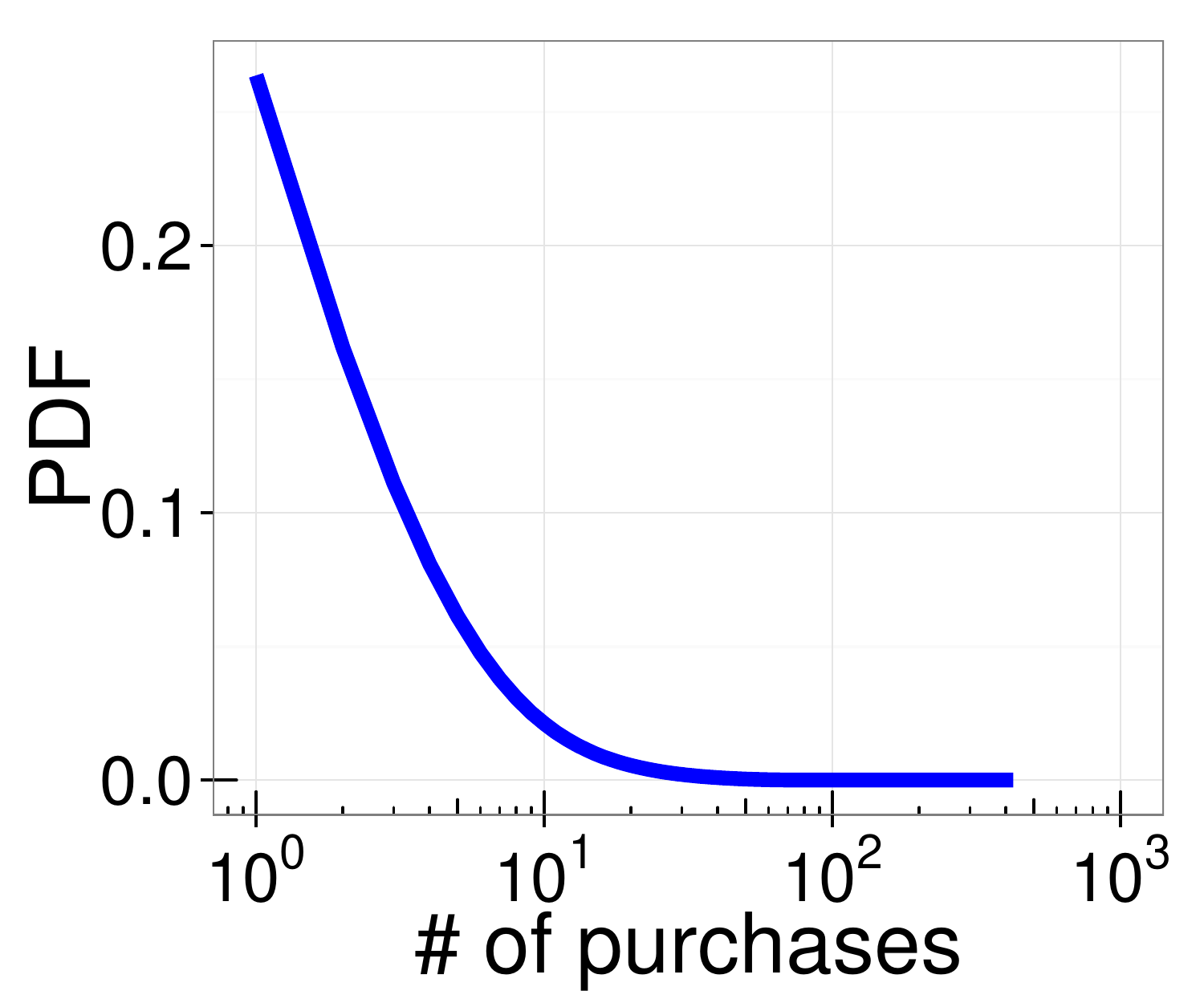}
      \label{fig:purchase_count_pdf}
   }
  &
   \subfigure[CDF]{
   \includegraphics[width=0.5\columnwidth]{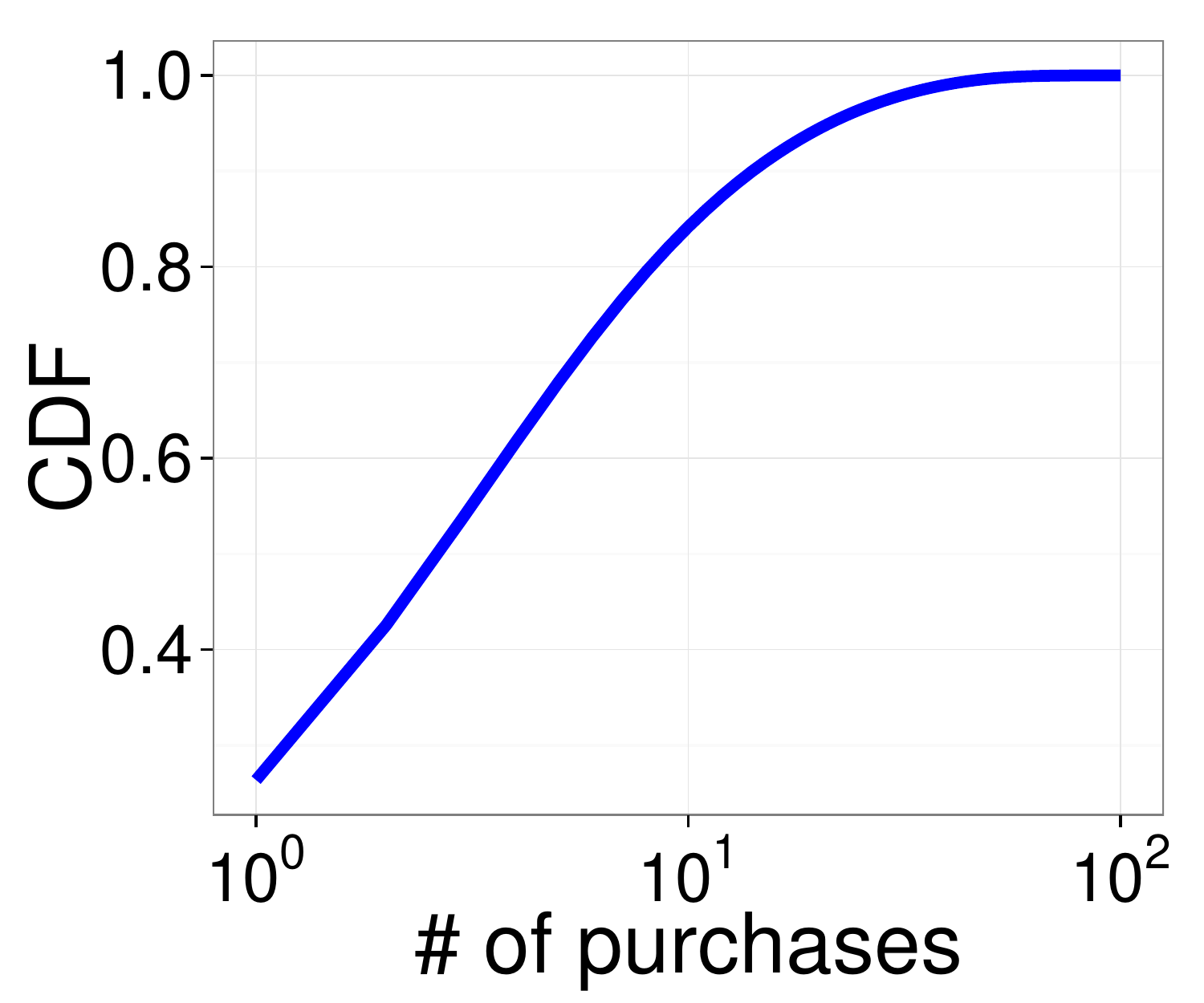}
   \label{fig:purchase_count_cdf}
   }
  \end{tabular}
   \caption{Distribution of number of purchases made by users.}
  \label{fig:purchase_count}
\end{figure}

Yahoo Mail is one of the world's largest email providers with more than 300M users\footnote{\url{http://www.comscore.com/}}, and many online shoppers use Yahoo Mail for receiving purchase confirmations. We select these emails by using a list of merchant's email addresses. Applying a set of manually written templates to the email body, we extract the list of purchased \textit{item names} and the \textit{price} of each item. The \textit{item name} was used as input to a classifier that predicts the \textit{item category}. We used a 3 levels deep, 1,733 node Pricegrabber taxonomy\footnote{\url{http://www.pricegrabber.com}} to categorize the items. The details of categorization are beyond the scope of this paper. In case of multiple items purchased in a single order, we consider them as individual purchases occurring at the same time. Therefore, throughout the paper the expression ``purchase" will refer to a purchase of a single item.

We limit our study to a random subset of Yahoo Mail users in the US; for all of them \textit{age}, \textit{gender}, and \textit{zip code} information is also available from the Yahoo user database. We excluded users who made more than 1,000 purchases (less than $0.01\%$ of the sample), because these accounts probably belong to stores and not to individuals. Overall, our dataset contains information on 20.1M users, who collectively made 121M purchases from several top retailers between February and September of 2014, amounting to total spending of 5.1B dollars. The dataset includes messages belonging exclusively to users who voluntarily opted-in for such studies. All the analysis has been performed in aggregate and on anonymized data. 

In addition, we use the Yahoo email network to be able to examine the social aspects of shopping behavior. The Yahoo email network is a directed graph $G$, where the edges are endowed with a list of times. We denote the edges $G$ by $(i, j, \{t_{ijk}\})$, which signifies that user $i$ emailed user $j$ at times $t_{ijk}$, $1 \le k \le N_{ij}$, where $N_{ij}$ is the total number of emails sent from user $i$ to user $j$. For the present analysis, we consider the subgraph of $G$ induced by the two-hop neighborhood from the users who made purchases $\mathcal{C}$, i.e., their immediate contacts and contacts of their contacts. The subgraph $\mathcal{C}$ was used to construct a list of 1st level contacts and 2nd level contacts for each online shopper. Only edges with a minimum of $5$ exchanged messages were retained.

\begin{figure}[t!]
\begin{tabular}{@{}c@{}c@{}}
\subfigure[PDF]{
   \includegraphics[width=0.5\columnwidth]{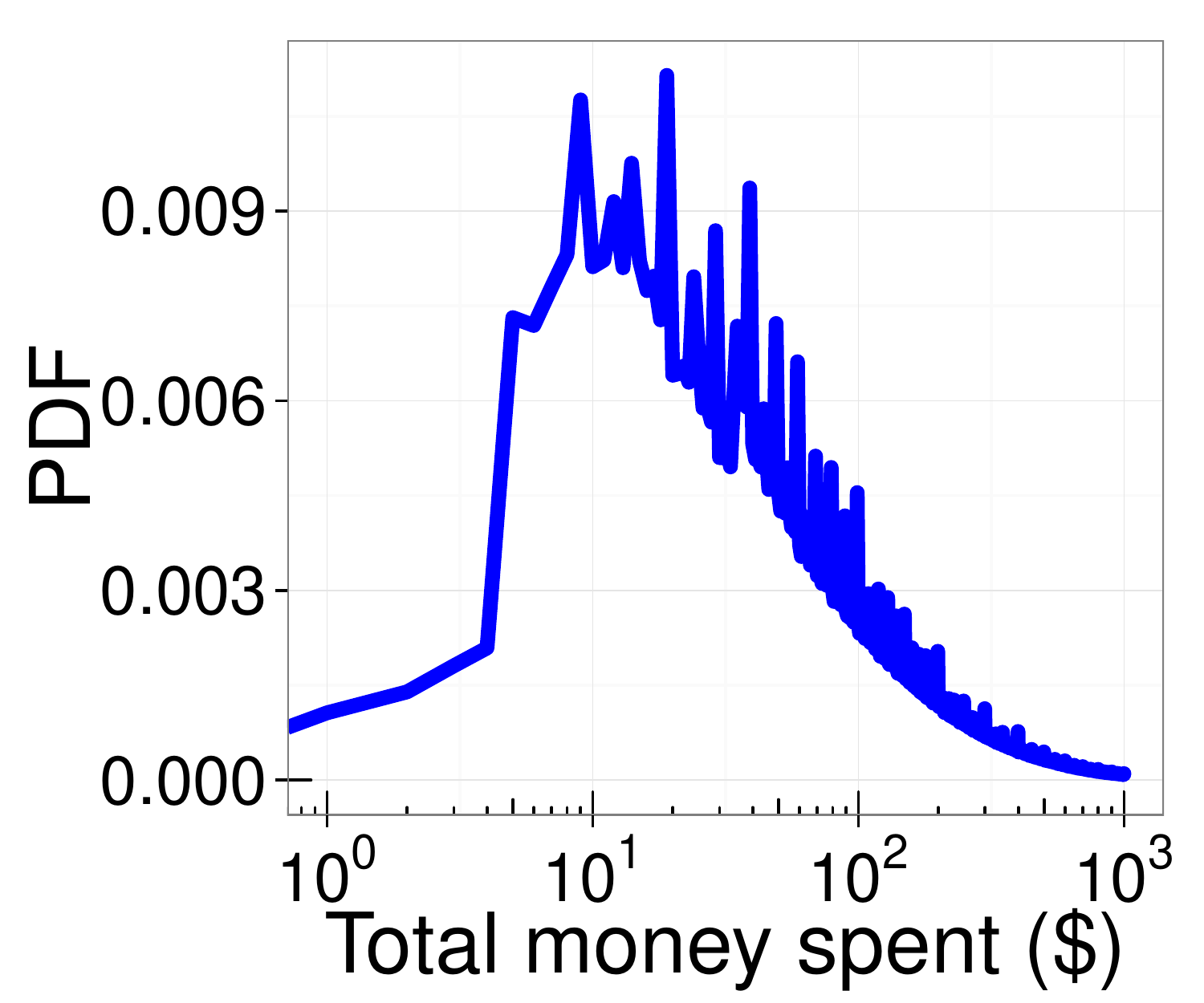}
      \label{fig:spent_pdf}
   }
  &
   \subfigure[CDF]{
   \includegraphics[width=0.5\columnwidth]{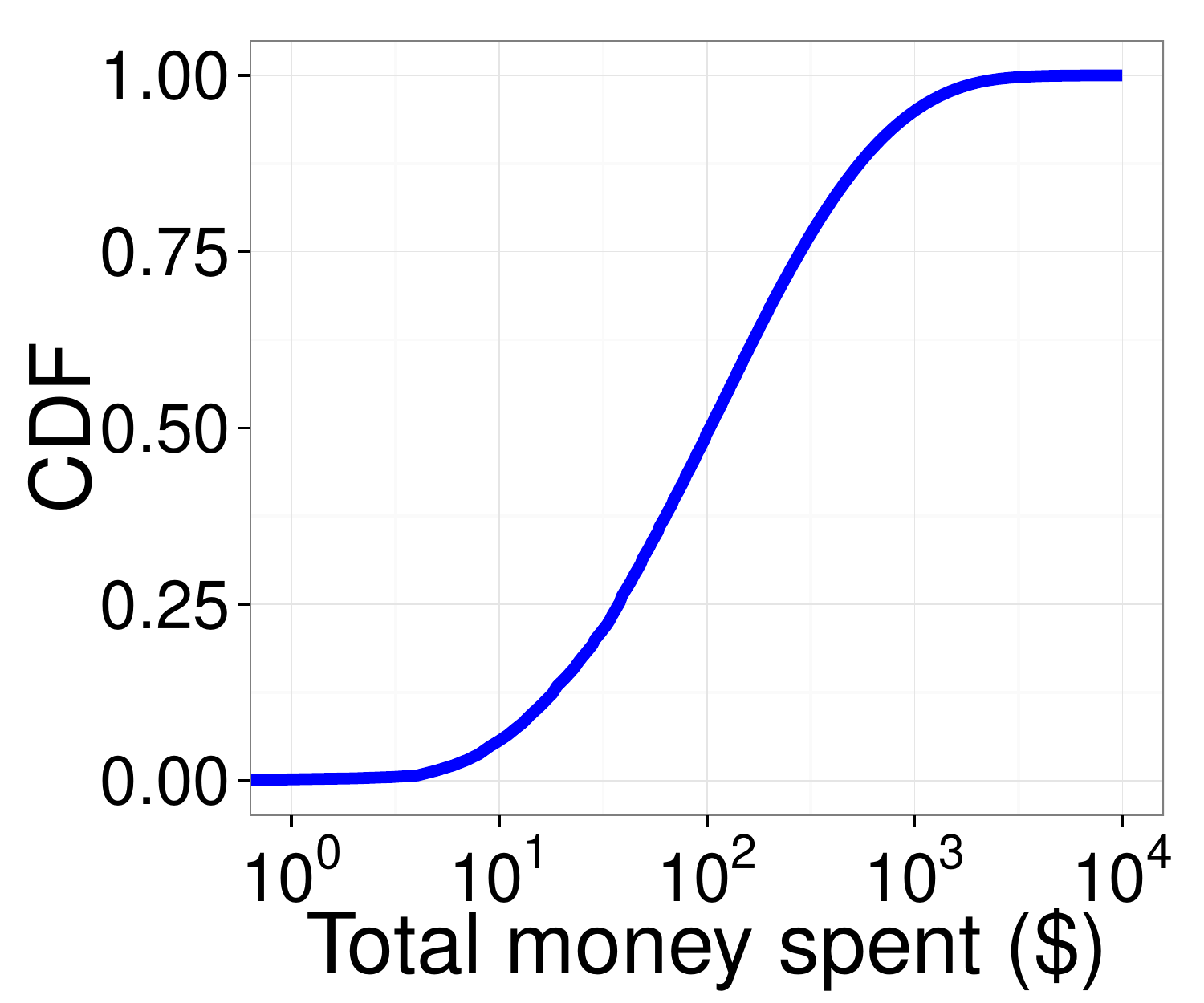}
   \label{fig:spent_cdf}
   }
  \end{tabular}
   \vspace{-2mm}
   \caption{Distribution of total money spent by users.}
  \label{fig:spent}
\end{figure}

\begin{figure}[t!]
\begin{center}
\includegraphics[width=0.8\columnwidth]{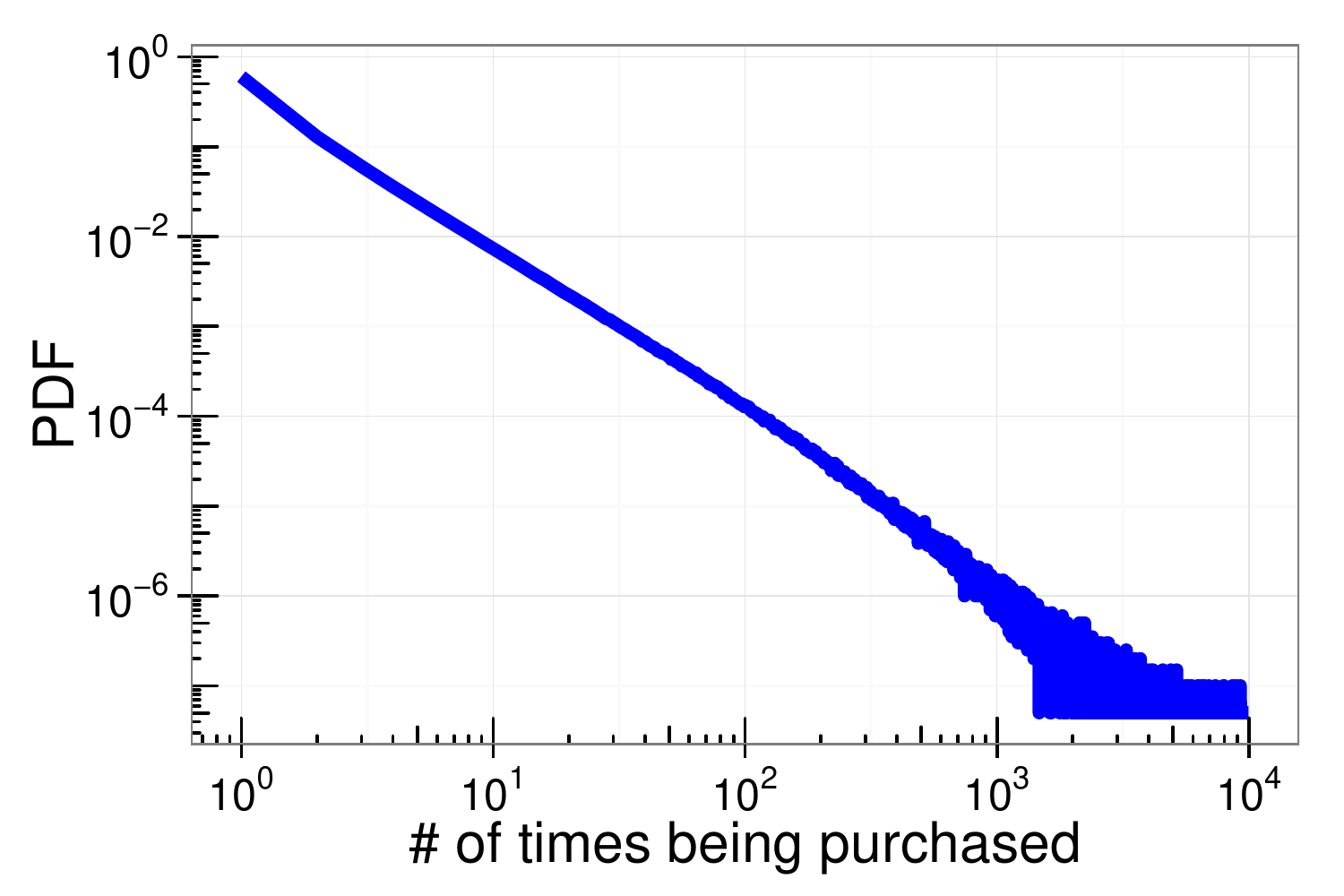}
\end{center}
\vspace{-2mm}
\caption{Distribution of number of times different items have been purchased.}
\label{fig:item_dist}
\end{figure}

As mentioned above, our dataset includes 121M purchases from 20.1M users. Figure~\ref{fig:purchase_count_pdf} shows the distribution of number of purchases made by users, showing the expected heavy-tailed characteristic. Figure~\ref{fig:purchase_count_cdf} shows that only 5\% of users made more than 20 purchases. In contrast, the distribution of total money spent peaks at around 10 dollars, sharply decreasing for smaller amounts (Figure~\ref{fig:spent_pdf}). Also, there is a non-negligible minority of people who spend a substantial amount of money for online shopping, e.g., 5\% of the users spent more than 1,000 dollars (Figure~\ref{fig:spent_cdf}).

\begin{figure}[t!]
\begin{center}
\includegraphics[width=0.8\columnwidth]{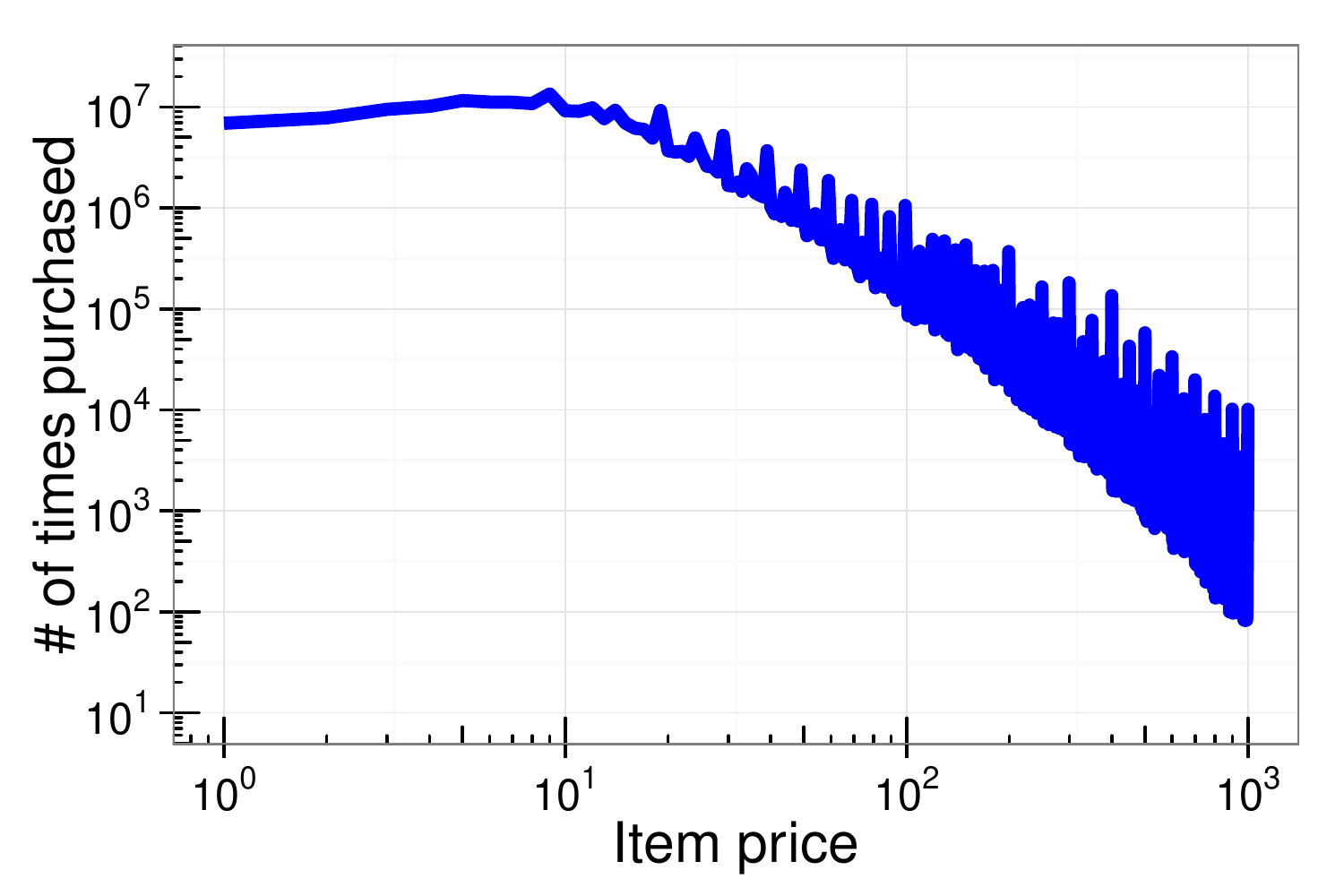}
\end{center}
\caption{Correlation of the price of the item and number of times being purchased.}
\vspace{-2mm}
\label{fig:purchase_price_dist}
\end{figure}

Items being purchased also have drastically different level of popularity, as shown by the distribution in Figure~\ref{fig:item_dist}. Disney's Frozen DVD, the most popular item in the dataset, has been sold more than 200,000 times, whereas the vast majority of items has been purchased less than 10 times. Table~\ref{table:top_items} lists the 5 most frequently purchased items. Intuitively, the set of items that users have spent the most money on is a different set, because a single purchase of an expensive item would account for the same amount of money of several purchases of cheaper items (Table~\ref{table:top_items_money}). In fact, the number of times an item is purchased negatively correlates with the price of that item (Figure~\ref{fig:purchase_price_dist}). This is in line with previous survey-based studies~\cite{bhatnagar00risk} that found that the vast majority of items purchased online are worth at most few tens of dollars, although in our dataset there is a consistent amount of shoppers who spend several thousands of dollars in repeated purchases over relatively short time periods.

\begin{table}[t!]
\begin {center}
{
\begin {tabular} {| c | l | r |}
\hline
{\textbf{Rank}} & {\textbf{Product name}} & {\textbf{\# of purchases}} \\
\hline
1 & Frozen (DVD) & 202,103\\
2 & Cards Against Humanity (Cards)& 110,032\\
3 & Google Chromecast &  59,548\\
4 & Amazon giftcard &  58,254\\
5 & HDMI cable & 54,402\\
\hline
\end{tabular}
}
\end{center}
\vspace{-1mm}
\caption{Top 5 most purchased Products.}
\label{table:top_items}
\end{table}

\begin{table}[t!]
\begin {center}
{
\begin {tabular} {| c | l | r |}
\hline
{\textbf{Rank}} & {\textbf{Product name}} & {\textbf{Money spent on product}} \\
\hline
1 & Play Station 4 & \$ 2.5M\\
2 & Amazon giftcard& \$ 2.0M\\
3 & Frozen (DVD) &  \$ 1.5M\\
4 & Fitbit fitness tracker &  \$ 1.4M\\
5 & Kindle & \$ 1.3M\\
\hline
\end{tabular}
}
\end{center}
\vspace{-1mm}
\caption{Top 5 products with the most money spent on them.}
\label{table:top_items_money}
\vspace{-2mm}
\end{table}

\section{Purchase Pattern Analysis}\label{section:behavior}


\noindent We present a quantitative analysis of factors affecting online purchases. We examine the role of demographic, temporal, and social factors including gender and age, daily and weekly patterns, frequency of shopping, tendency to recurrent purchases, and budget constraints.

\subsection{Demographic Factors}

\begin{figure*}[thb!]
\begin{center}
\subfigure[Percentage of online shoppers] {
\label{fig:purchase_demo_perc}
\includegraphics[width=0.23\textwidth]{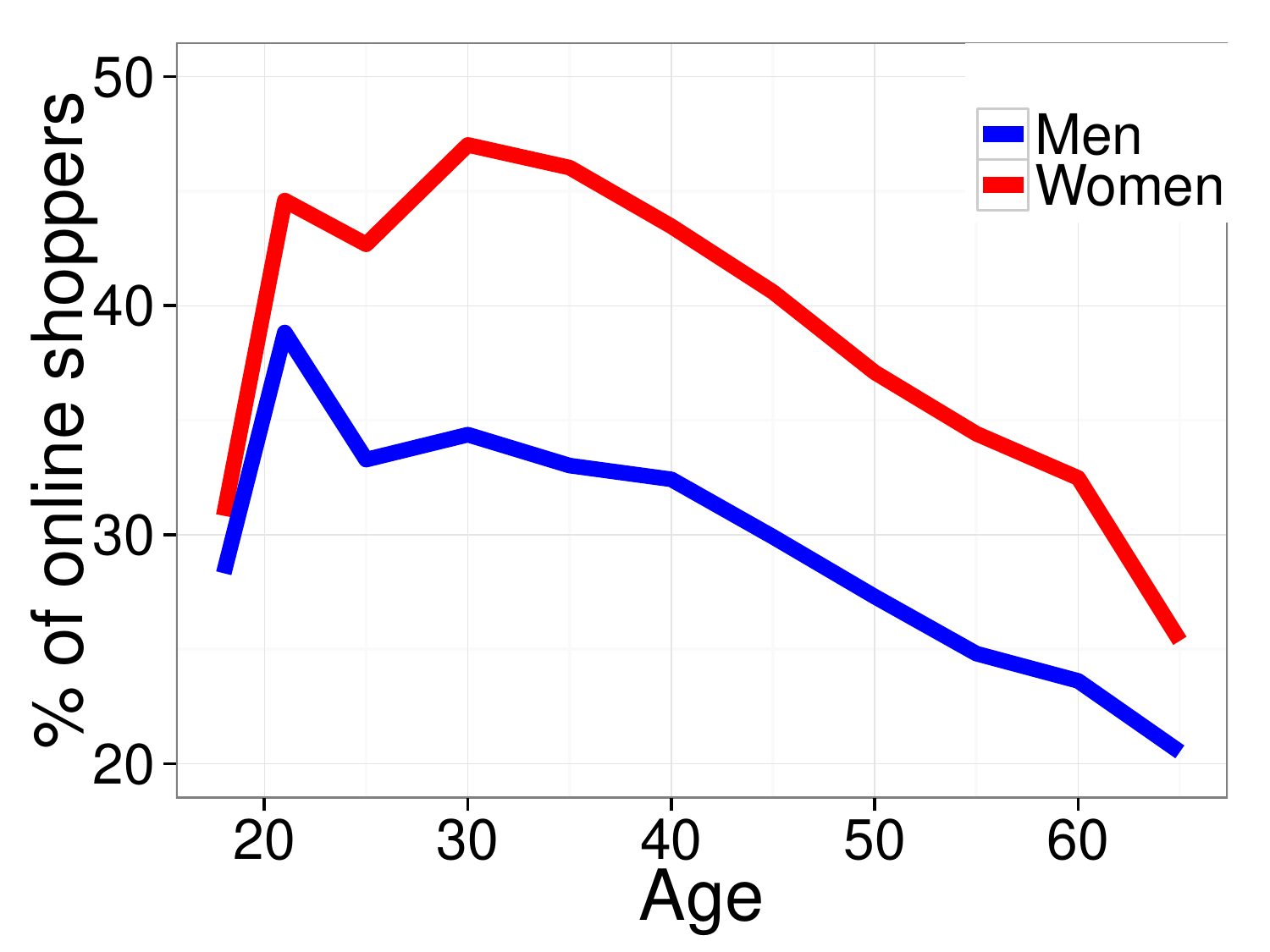}
}
\subfigure[Number of purchases] {
\label{fig:demo_count}
\includegraphics[width=0.23\textwidth]{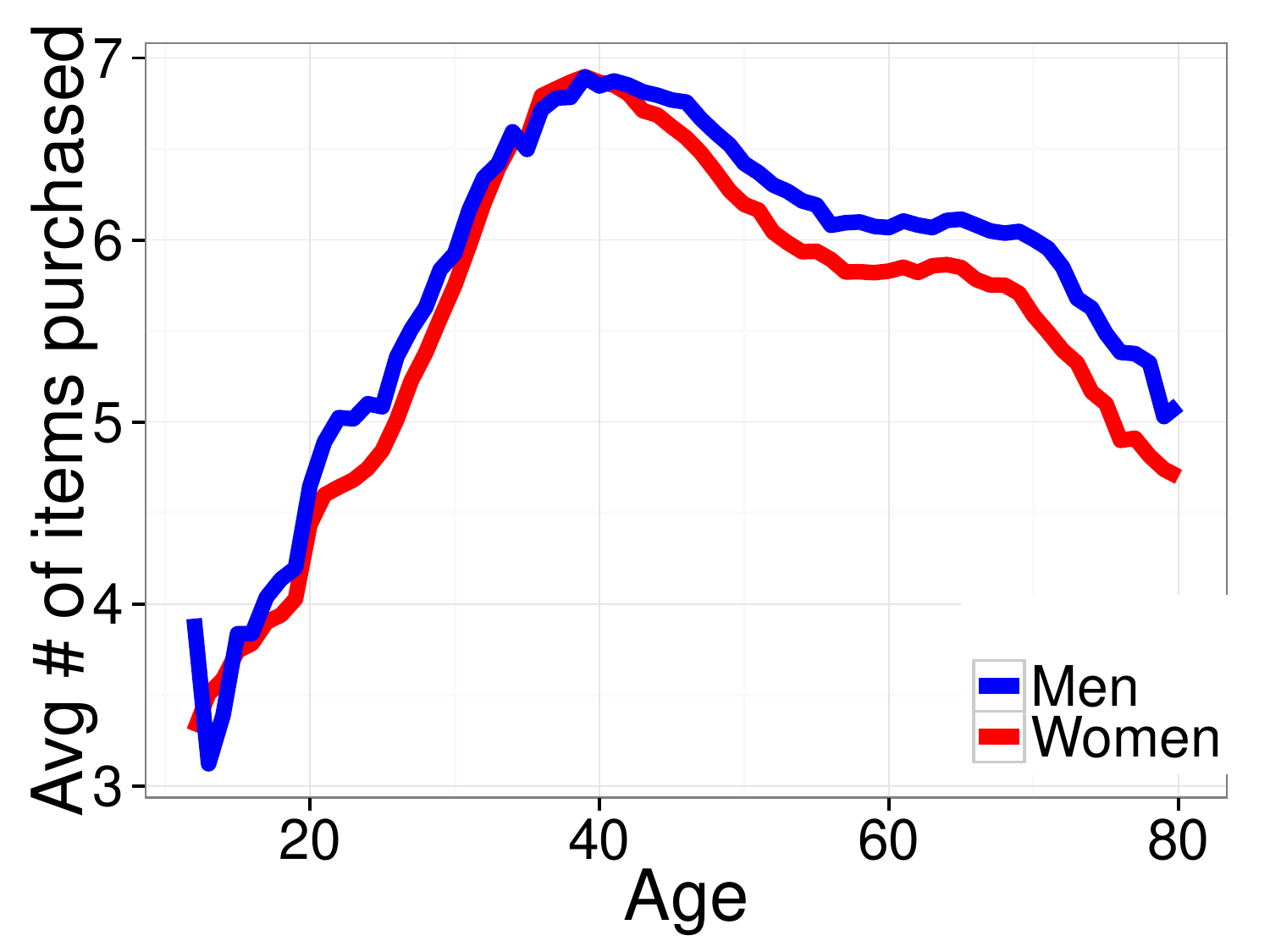}
}
\subfigure[Average price] {
\label{fig:demo_avg}
\includegraphics[width=0.23\textwidth]{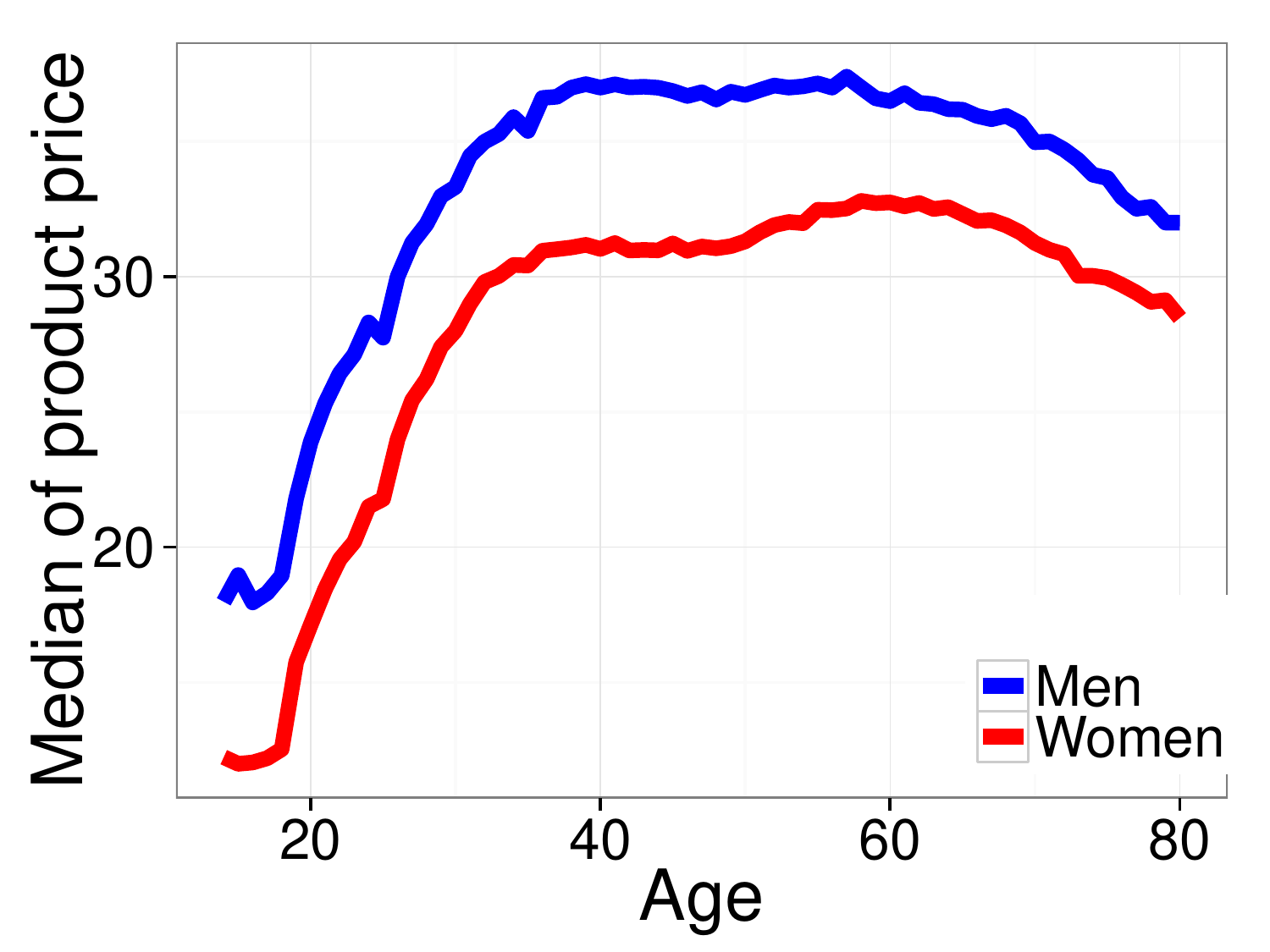}
}
\subfigure[Total money spent] {
\label{fig:demo_sum}
\includegraphics[width=0.23\textwidth]{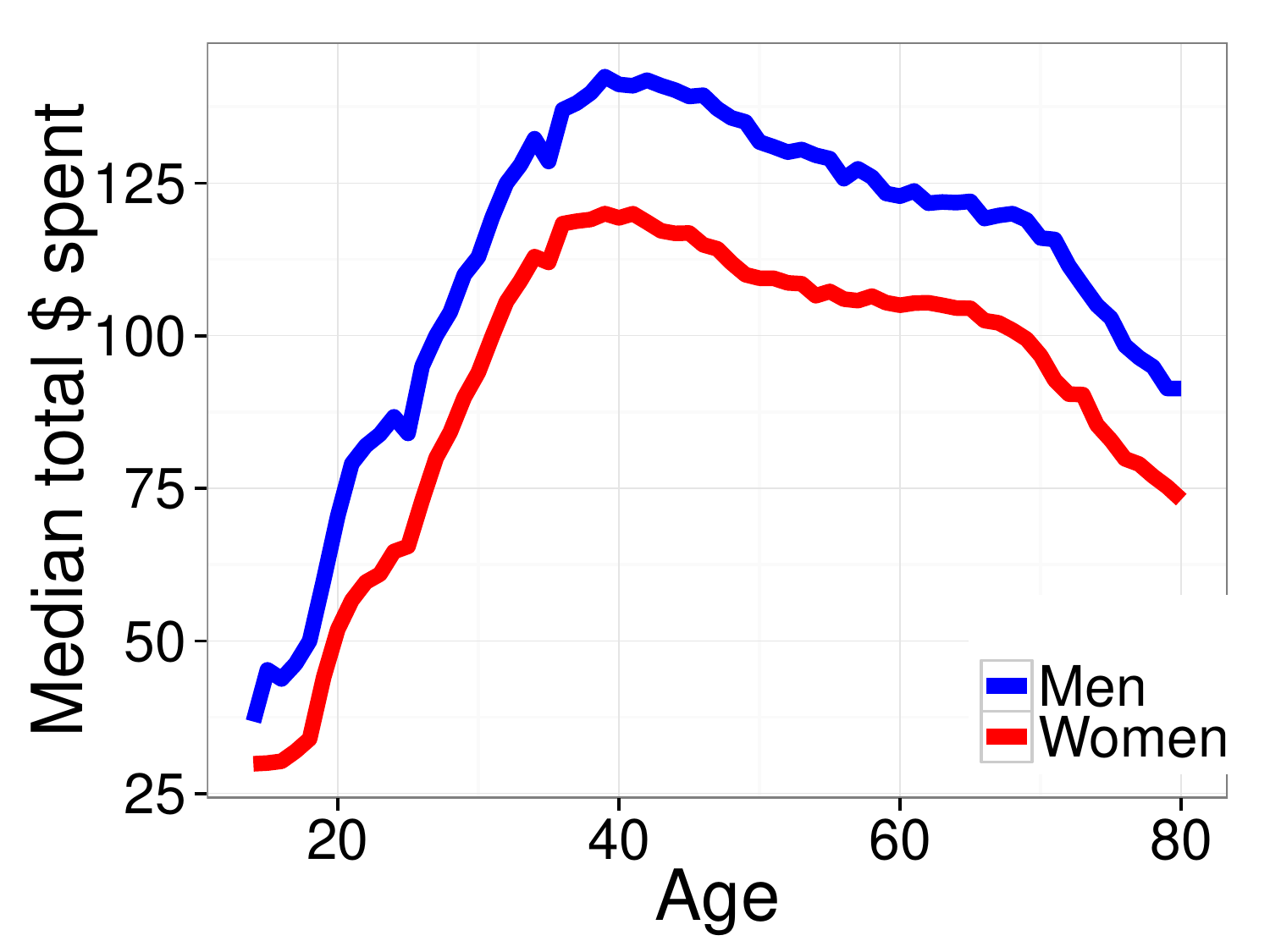}
}
\caption{Demographic analysis. (a) Percentage of online shoppers, (b) number of items purchased, (c) average price of products purchased, and (d) total spent by men and women, broken down by age.}
\label{fig:income}
\end{center}
\end{figure*}

We measure how gender, age, and location (zip code) affect purchasing behavior. First, we measure the fraction of all email users that made an online purchase. We find that higher fraction of women make online purchases compared to men ((Figure~\ref{fig:purchase_demo_perc})), but men make slightly more purchases per person (Figure~\ref{fig:demo_count}), and they spend more money, on average, on online purchases (Figure~\ref{fig:demo_avg}). As a result, men spend much more money in total (Figure~\ref{fig:demo_sum}). The same patterns hold across different age groups. All these plots, except Figure~\ref{fig:purchase_demo_perc} back up findings from earlier consumer surveys that revealed man having a higher perceived advantage of online shopping~\cite{vanslyke05gender}, and women having higher concern of negative consequences of online purchasing~\cite{garbarino04gender}, resulting in a higher number of purchases done by men.

\begin{table}[t!]
\begin {center}
{
\begin {tabular} {| c | l | l | l | l |}
\hline
\textbf{Rank} & \textbf{Top categories} & \textbf{Distinctive women} & \textbf{Distinctive men}  \\
\hline
1 & Android & Books &  Games\\
2 & Accessories & Dresses & Flash memory\\
3 & Books & Diapering & Light bulbs\\
4 & Vitamins & Wallets & Accessories\\
5 & Shirts & Bracelets & Batteries\\
\hline
\end{tabular}
}
\end{center}
\vspace{-1mm}
\caption{Differences in the categories of products purchased by women and men.}
\label{tab:gender_category}
\vspace{-3mm}
\end{table}

With respect to the age, spending ability increases as people get older, peaking among the population between age 30 to 50 and declining afterwards. The same pattern holds for number of purchases made, average item price, and total money spent (Figures~\ref{fig:demo_count},~\ref{fig:demo_sum},~\ref{fig:demo_avg}). 

Men and women also purchase different types of products online. Table~\ref{tab:gender_category} shows the top five categories of purchased products. Even though the ranking of the top products is the same for men and women, each product accounts for different fraction of all purchases within the same gender. To find the most distinctive categories, we compare the fraction of all the items bought by both genders, and consider the categories that have the largest differences. Books, dresses, and diapering are the categories that are more disproportionately bought by women, whereas games, flash memory sticks, and accessories (e.g., headphones) are the categories purchased more by men. The largest differences range from only 0.5\% to 1\%, but are statistically significant. This result is aligned with previous research on offline shopping that found men more keen in buying electronics and entertainment products and women more inclined to buy clothes~\cite{dholakia99going,JOCA:JOCA113}. We repeat the same gender analysis at product level (Table~\ref{tab:gender_item}). Consistently, there is high overlap between most purchased products, but no overlap between the most gender-distinctive ones. Differences exist across ages also (Table~\ref{tab:age_item}). Younger shoppers (18-22 years old) purchase more phone accessories and games, whereas older shoppers (60-70 years old) are much more interested in buying TV shows. Also, blood sugar medicine is purchased more by the older users, which is expected.



\begin{table*}[t!]
\begin {center}
{
\begin {tabular} {| c | l | l | l | l | l |}
\hline
\textbf{Rank} & \textbf{Top women} & \textbf{Top men} & \textbf{Distinctive women} & \textbf{Distinctive men}  \\
\hline
1 & Frozen & Frozen & Frozen & Chromecast \\
2 & iPhone screen protector & Game of Thrones & iPhone screen protector & Game of Thrones\\
3 & Amazon Giftcard & Chromecast & Amazon Giftcard & Titanfall Xbox One\\
4 & Cards Against Humanity & Cards Against Humanity & iPhone screen protector & Playstation 4\\
5 & iPhone screen protector (another brand) & Amazon gift card & Eyelashes & Godfather collection\\
\hline
\end{tabular}
}
\end{center}
\caption{Differences in the products purchased by women and men.}
\label{tab:gender_item}
\end{table*}

\begin{table*}[t!]
\begin {center}
{
\begin {tabular} {| c | l | l | l | l | l |}
\hline
\textbf{Rank} & \textbf{Top younger users} & \textbf{Top older users} & \textbf{Distinctive younger users} & \textbf{Distinctive older users}  \\
\hline
1 & iPhone screen protector & Frozen & iPhone screen protector & Frozen\\
2 & iPhone screen protector & Amazon gift card & iPhone screen protector (another brand) &  Amazon gift card\\
3 & Cards Against Humanity & Game of Thrones & Cards Against Humanity & Game of Thrones\\
4 & iPhone case & Chromecast & iPhone case & Downton Abbey\\
5 & Frozen & Downton Abbey & iPhone case &  Blood sugar medicine\\
\hline
\end{tabular}
}
\end{center}
\caption{Differences in the products purchased by younger (18-22 yo) and older (60-70 yo) users.}
\label{tab:age_item}
\end{table*}

\begin{figure*}[thb!]
\begin{center}
\subfigure[Number of purchases] {
\label{fig:income_count}
\includegraphics[width=0.3\textwidth]{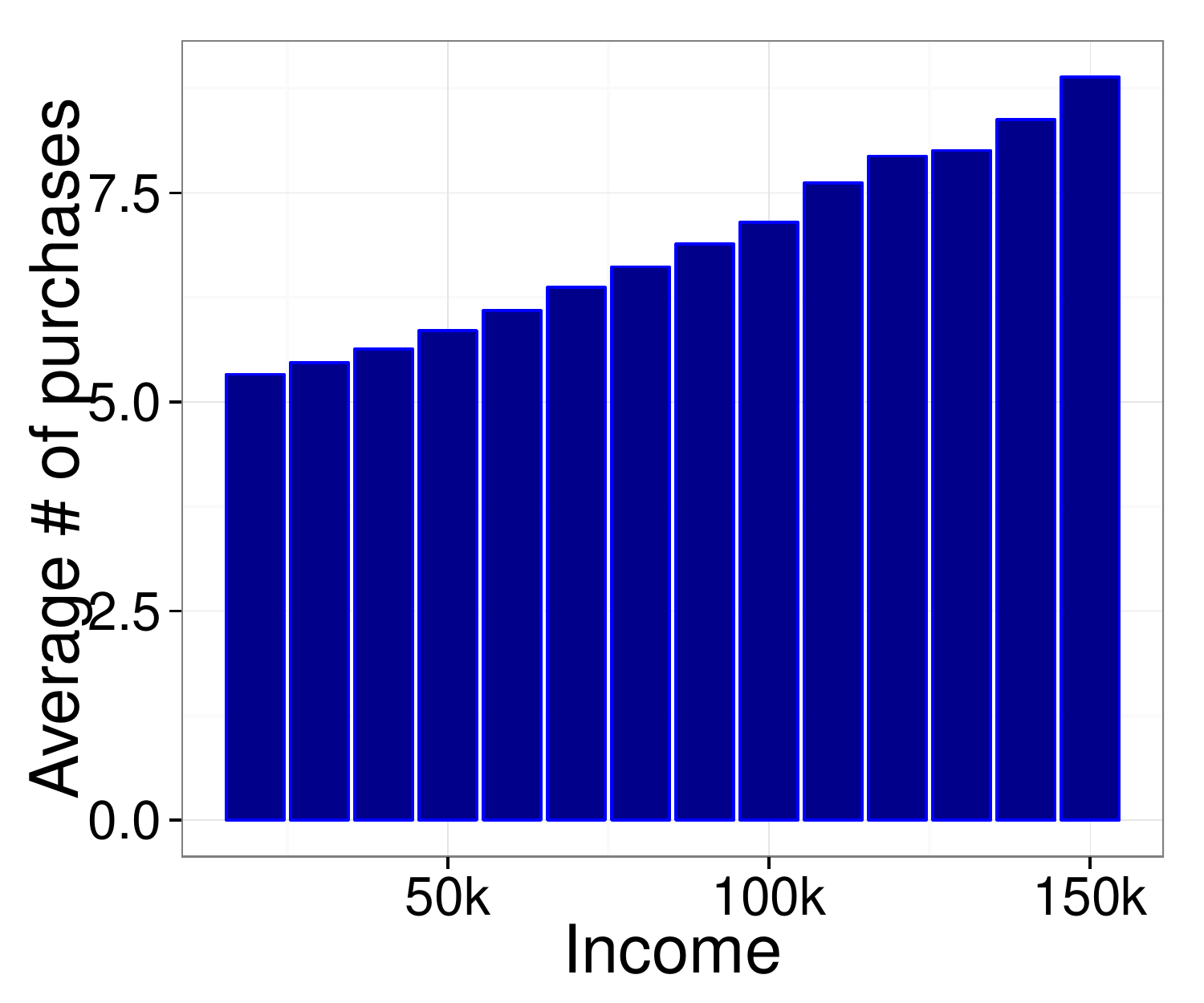}
}
\subfigure[Average item price] {
\label{fig:income_avg}
\includegraphics[width=0.3\textwidth]{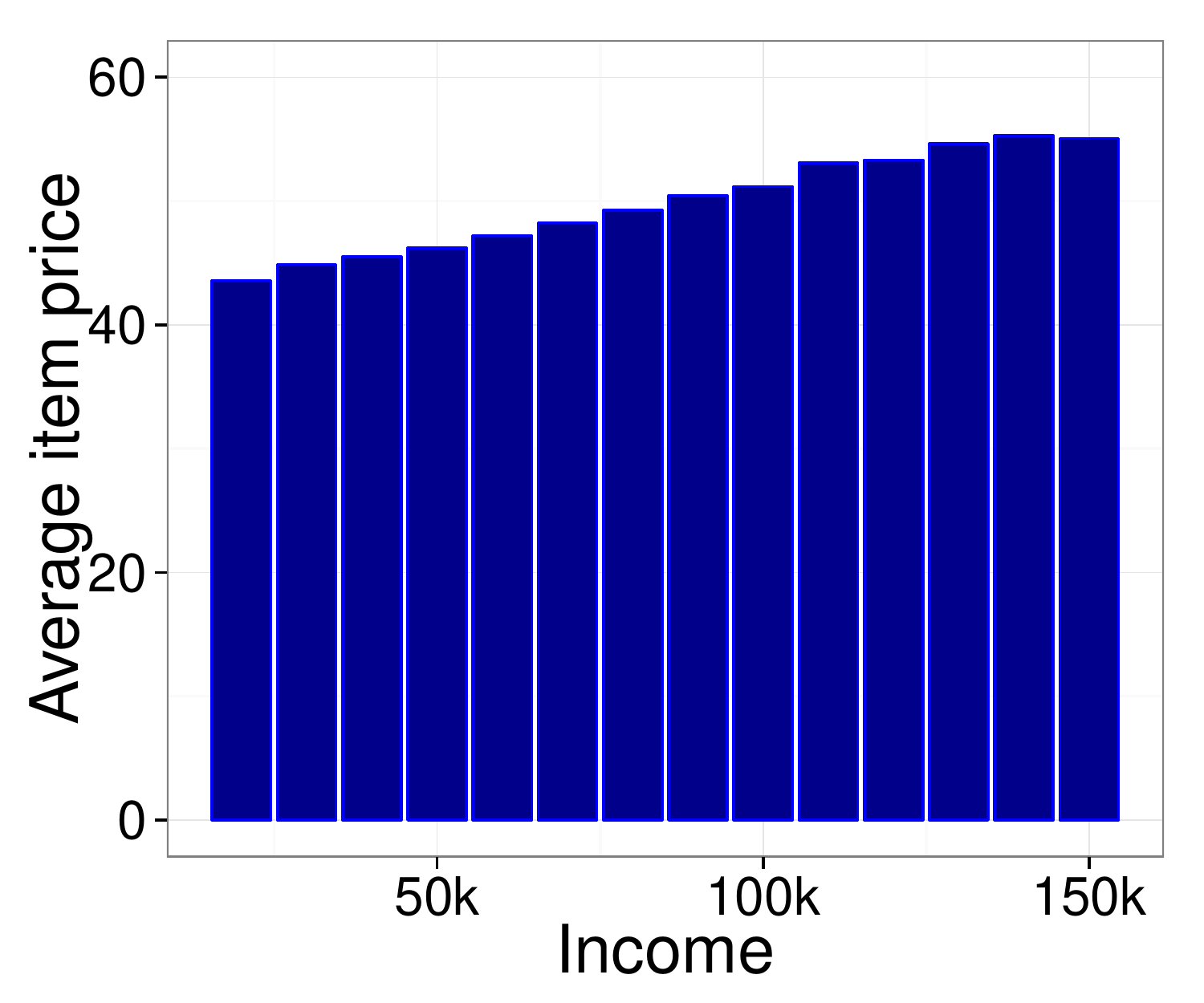}
}
\subfigure[Total money spent] {
\label{fig:income_sum}
\includegraphics[width=0.3\textwidth]{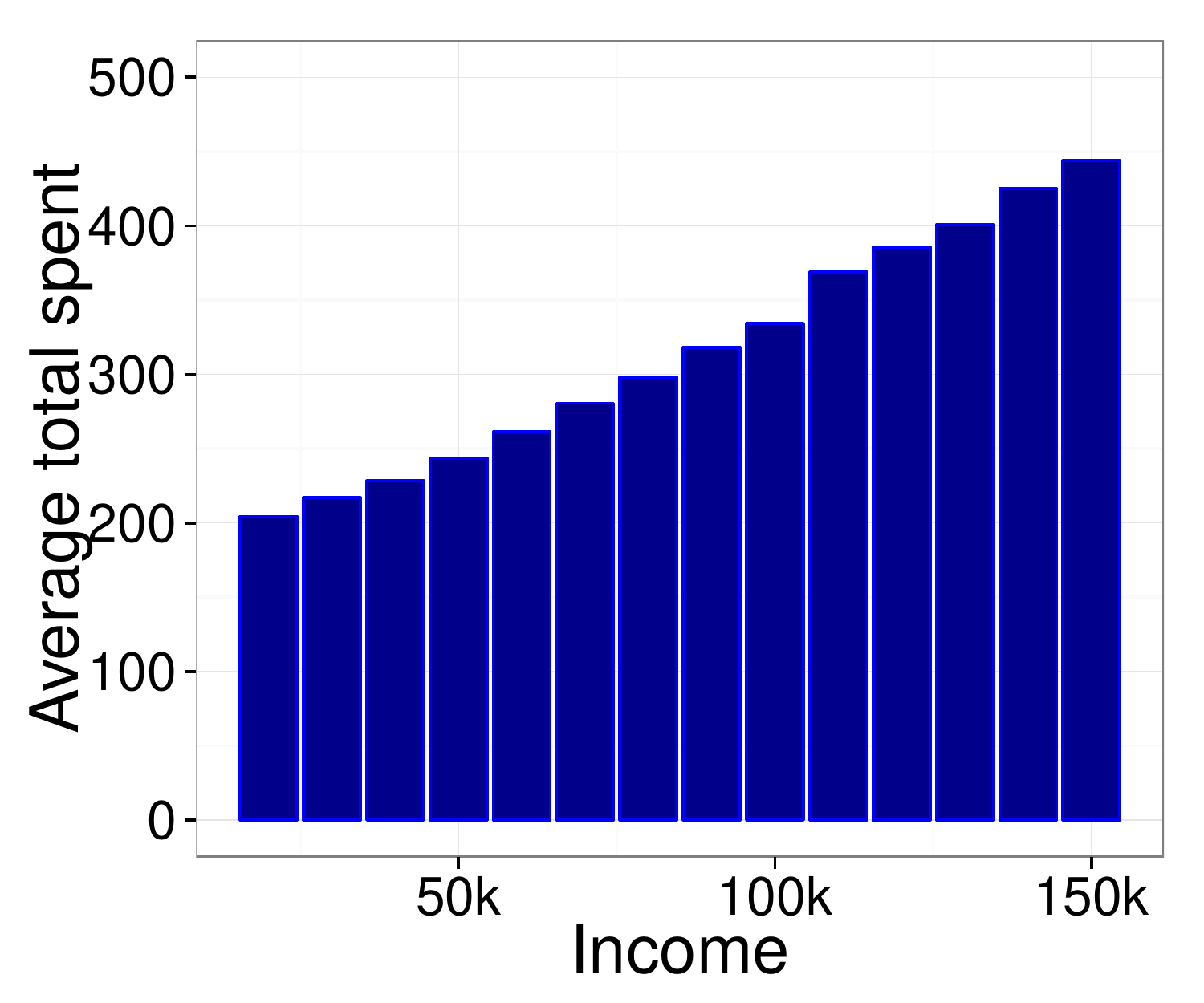}
}
\caption{Effect of income on purchasing behavior.}
\label{fig:income}
\end{center}
\end{figure*}

We also measure the impact of economic factors on online shopping behavior. We use US Census data to retrieve the median income associated with each zip code\footnote{\url{http://www.boutell.com/zipcodes/zipcode.zip}}. The inferred income for the user is an aggregated estimate but, given the large size of this dataset, this coarse appraisal is enough to observe clear trends. The number of purchases, average product price, and total money spent (Figures~\ref{fig:income_count},~\ref{fig:income_avg}, and ~\ref{fig:income_sum} respectively) are all positively correlated with income. While users living in high income zip codes do not buy substantially more expensive products, they make many more purchases, spending more money in total than users from lower income zip codes. Although the factors leading lower-income households to spend less online are multiple and entangled, part of this effect can be explained by the reluctance of people who are concerned with their financial safety to trust and make full use of online shopping, as it has been pointed out by previous studies~\cite{horrigan2008online}.

\subsection{Temporal Factors}
Our dataset spans a period of eight months, giving us opportunity to investigate temporal dynamics of purchasing behavior and factors affecting it. Besides daily and weekly cycles and periodic purchasing, we observe temporal variations that we associate with financial depletion: users wait longer to buy more expensive items, waiting for the budget to recover from the previous purchases.

\newpage

\subsubsection{Daily and Weekly Cycles}
\begin{figure}[t!]
\begin{center}
\includegraphics[width=0.9\columnwidth]{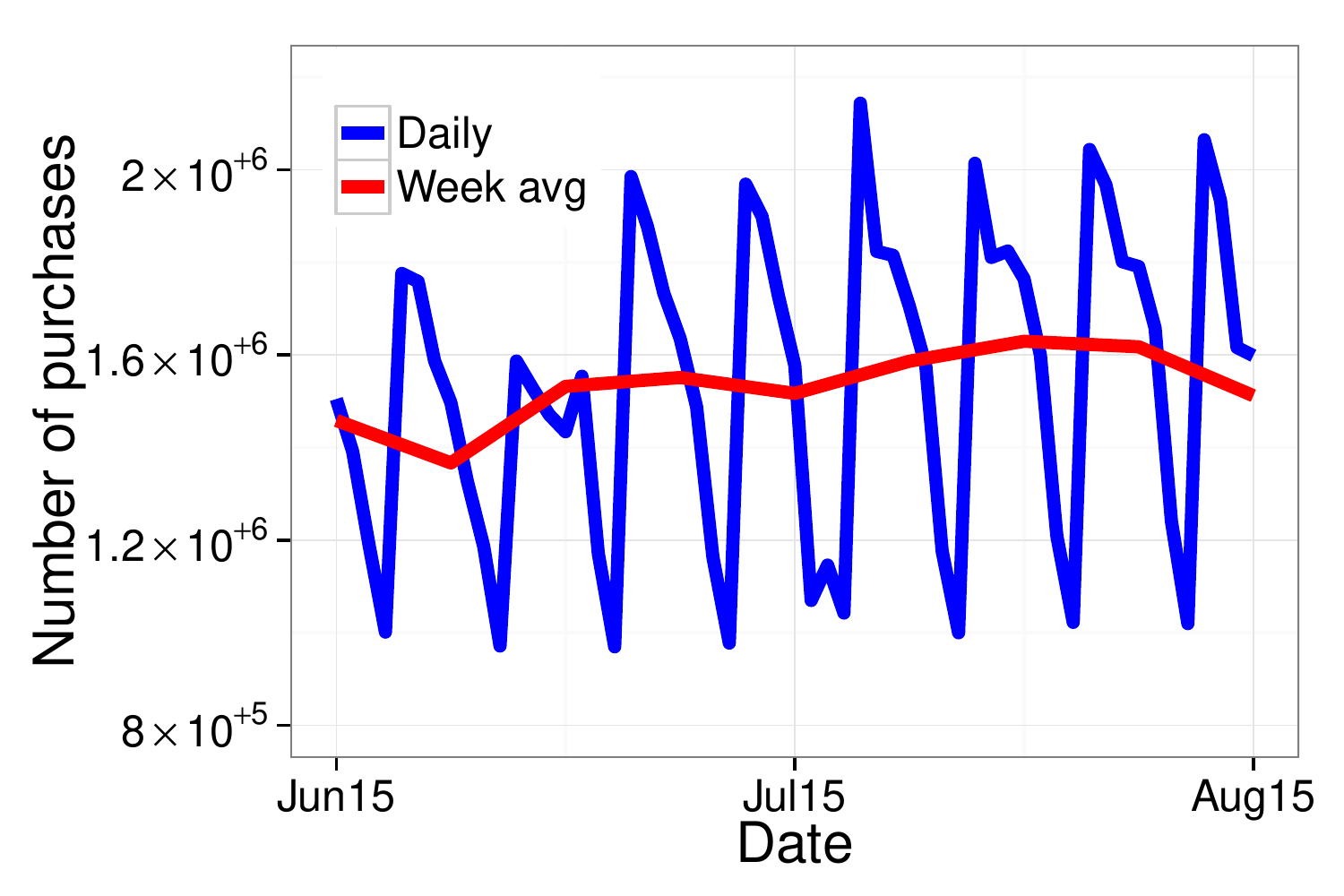}
\end{center}
\vspace{-1mm}
\caption{Number of purchases in a day and average weekly
\vspace{-2mm}. 
}
\label{fig:purchase_time_dist}
\end{figure}

Figure~\ref{fig:purchase_time_dist} shows the daily number of purchases over a period of two months. The figure shows a clear weekly shopping pattern with more purchases taking place in the first days of the week and fewer purchases on the weekends. On average there are 32.6\% more purchases on Mondays than Sundays.

\begin{figure}[t!]
\begin{center}
\includegraphics[width=0.9\columnwidth]{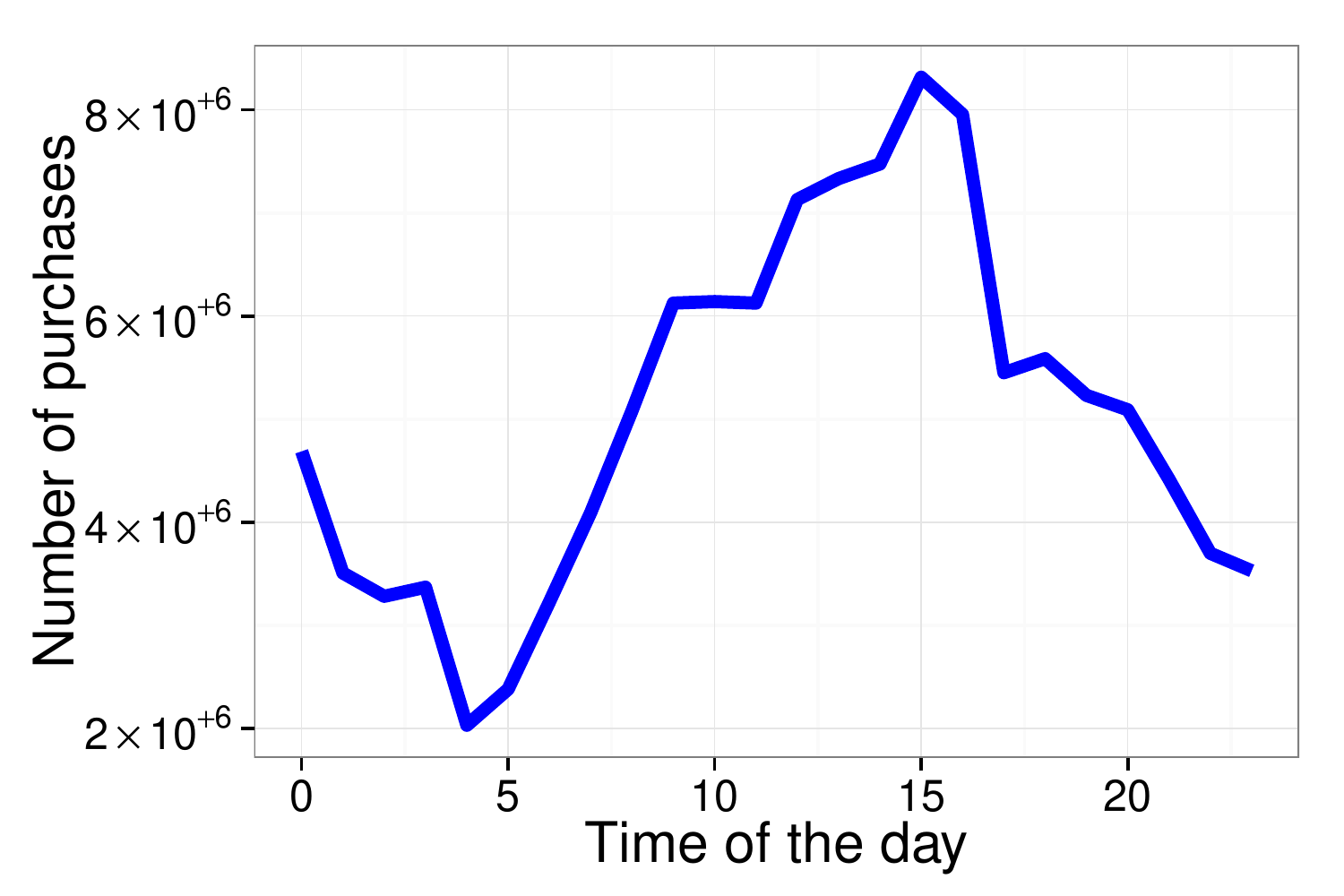}
\end{center}
\vspace{-1mm}
\caption{Number of purchases in each hour of the day.}
\vspace{-2mm}
\label{fig:diurnal}
\end{figure}

There is also strong diurnal pattern in shopping (Figure~\ref{fig:diurnal}). Interestingly, most of the purchases occur during the working hours, i.e. in the morning and early afternoon. However, note that for this analysis we infer the time zone from the user's zip code, which might be different from a shipping zip code for a purchase.
 Researchers have also reported monthly effects, where people spend more money at the beginning of the month when they receive their paychecks, compared to the end of the month~\cite{Evans11}. To test the \textit{first-of-the-month} phenomenon, we compared spending in the first Monday of the month with the last Monday of the month. We considered the first and the last Mondays and not the first and the last days, because the strong weekly patterns would result in an unfair comparison if the first and the last day of the month are not the same day of the week. Our data does not support the earlier findings and there are months in which the last Monday of month includes more activity compared to the first Monday of the month.

\subsubsection{Recurring Purchases}

\begin{table}[t!]
\begin {center}
{
\begin {tabular} {| c | l | r |}
\hline
{\textbf{Rank}} & {\textbf{Item name}} & {\textbf{Median purchase delay}} \\
\hline
1 & Pampers 448 count & 42\\
2 & Bath tissue & 62\\
3 & Pampers 162 count &  30\\
4 & Pampers 152 count &  31\\
5 & Frozen & 12\\
\hline
\end{tabular}
}
\end{center}
\caption{Top 5 items with the most number of repurchases.}
\vspace{-2mm}
\label{table:top_repurchase}
\end{table}

Some products are purchased periodically, such as printer cartridges, water filters, and toilet paper. Finding these items and their typical cycle would help predicting purchasing behavior. We do this by counting the number of times each item has been purchased by each user, then from each user's count we eliminate those products purchased only once, and last we aggregate the number of purchases per each item. Table~\ref{table:top_repurchase} shows top five such products, along with the number of times they were purchased, and the median number of days between purchases. Out of the top 20 products only four are neither toilet paper nor tissue (Frozen, Amazon gift card, chocolate chip cookie dough, and single serve coffees). In the top 20 list, the only unexpected item is the Frozen DVD, which probably made the list due to users buying additional copies as gifts or due to purchases by small stores that were not eliminated by our maximum 1,000 purchases removal criteria. Interestingly, the number of days between purchases for most of the top 20 items is close to 1 or 2 months, which might be due to automatic purchasing that users can set up.



\subsubsection{Finite Budget Effects}

\begin{figure}[t!]
\begin{center}
\includegraphics[width=0.8\columnwidth]{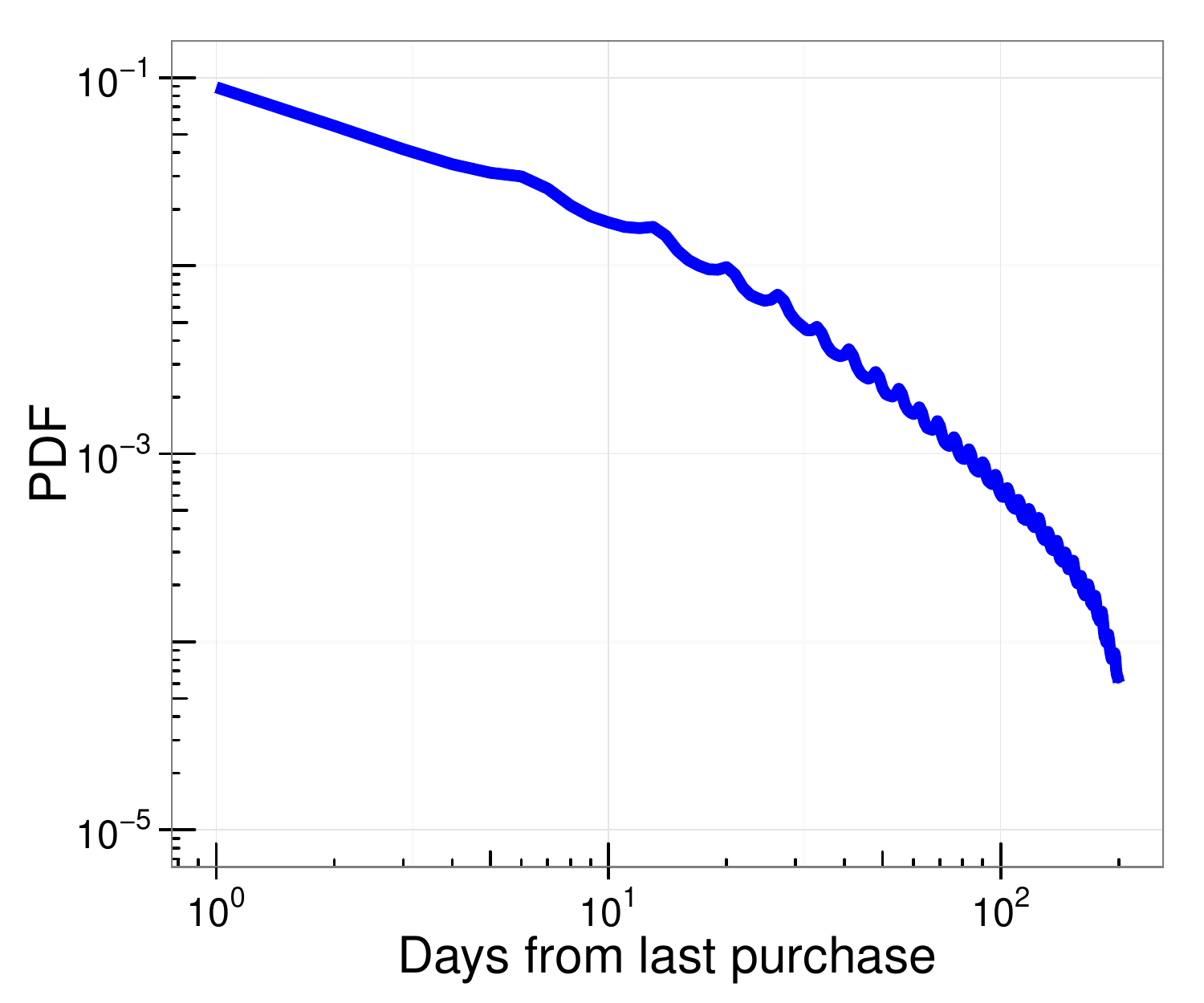}
\end{center}
\caption{Distribution of number of days between purchases.}
\label{fig:day_dist_pdf}
\end{figure}

Finally, we study the dynamics of individual purchasing behavior.
Figure~\ref{fig:day_dist_pdf} shows the distribution of number of days between purchases. The distribution is heavy-tailed, indicating bursty dynamics. The most likely time between purchases is  one day and there are local maxima around multiples of 7 days, consistent with weekly cycles we observed. 

\begin{figure}[t!]
\begin{center}
\includegraphics[width=0.8\columnwidth]{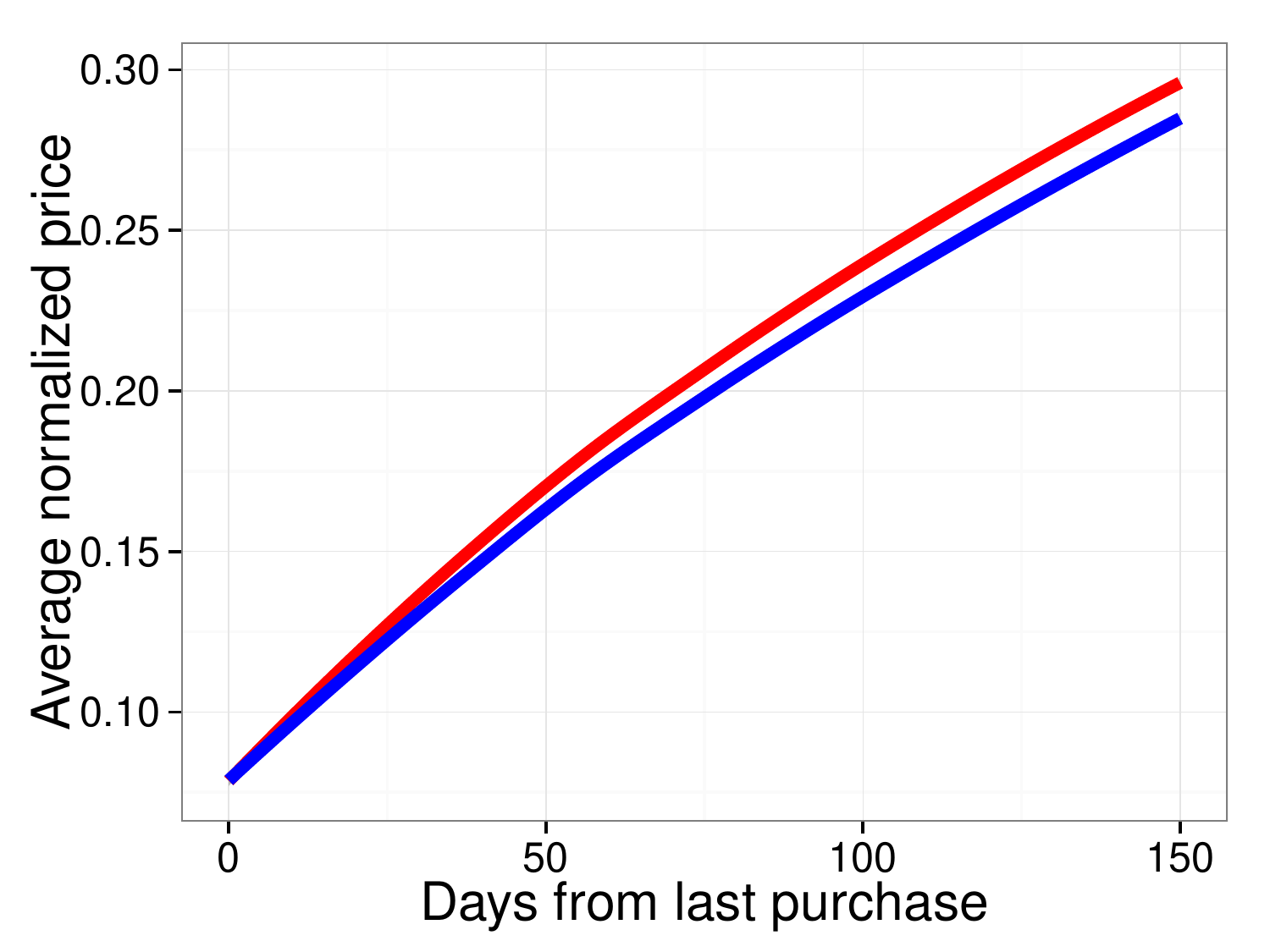}
\end{center}
\caption{Relationship between purchase price and time to next purchase. 0.95 confidence interval is also shown, but it is too small to be observed.}
\vspace{-2mm}
\label{fig:last_one}
\vspace{-2mm}
\end{figure}

\begin{figure*}[thb!]
\begin{center}
\subfigure[Users with 5 purchases] {
\label{fig:last_5}
\includegraphics[width=0.32\textwidth]{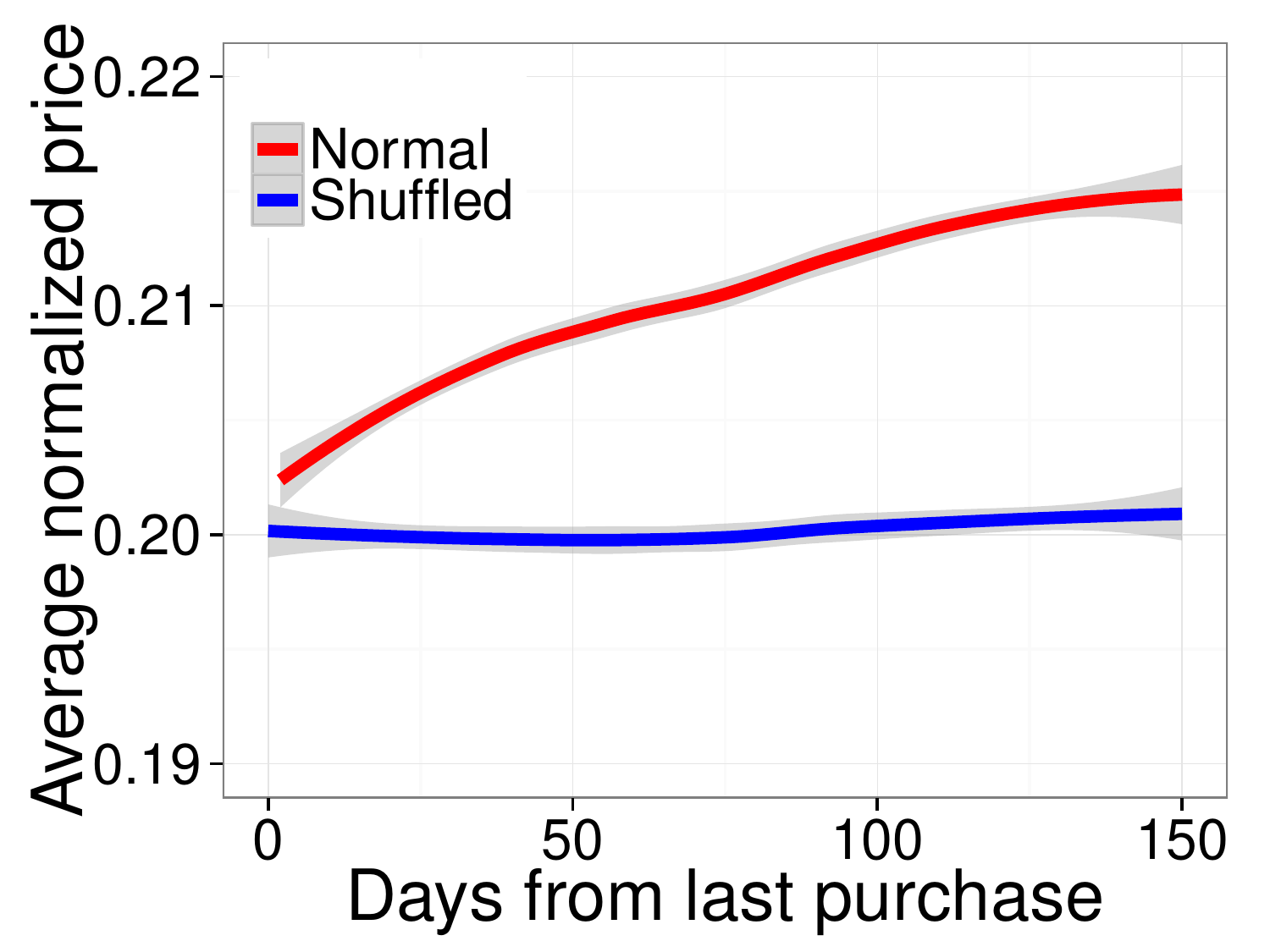}
}
\subfigure[Users with 9-11 purchases] {
\label{fig:last_9_11}
\includegraphics[width=0.32\textwidth]{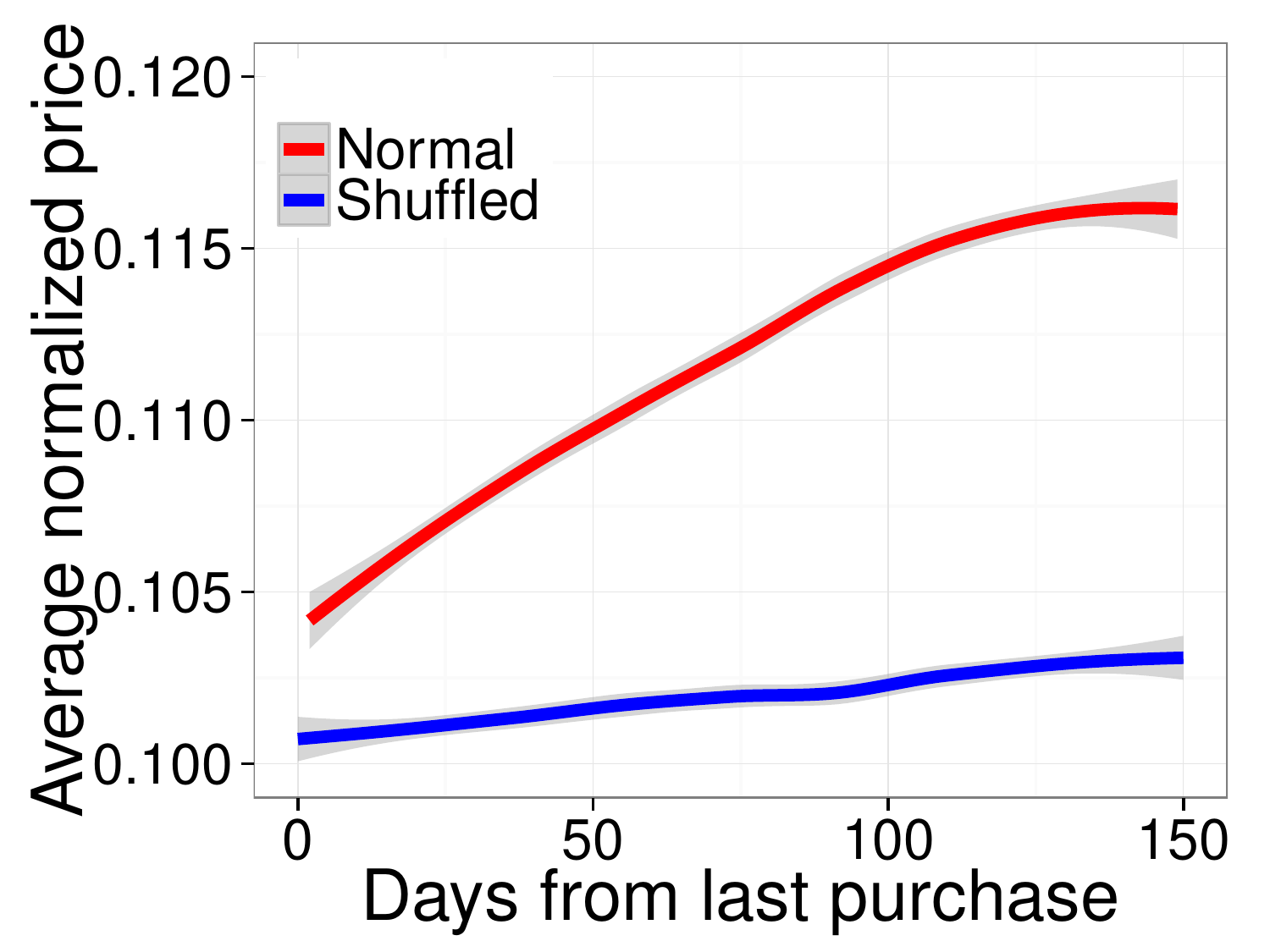}
}
\subfigure[Users with 28-32 purchases] {
\label{fig:last_28_32}
\includegraphics[width=0.32\textwidth]{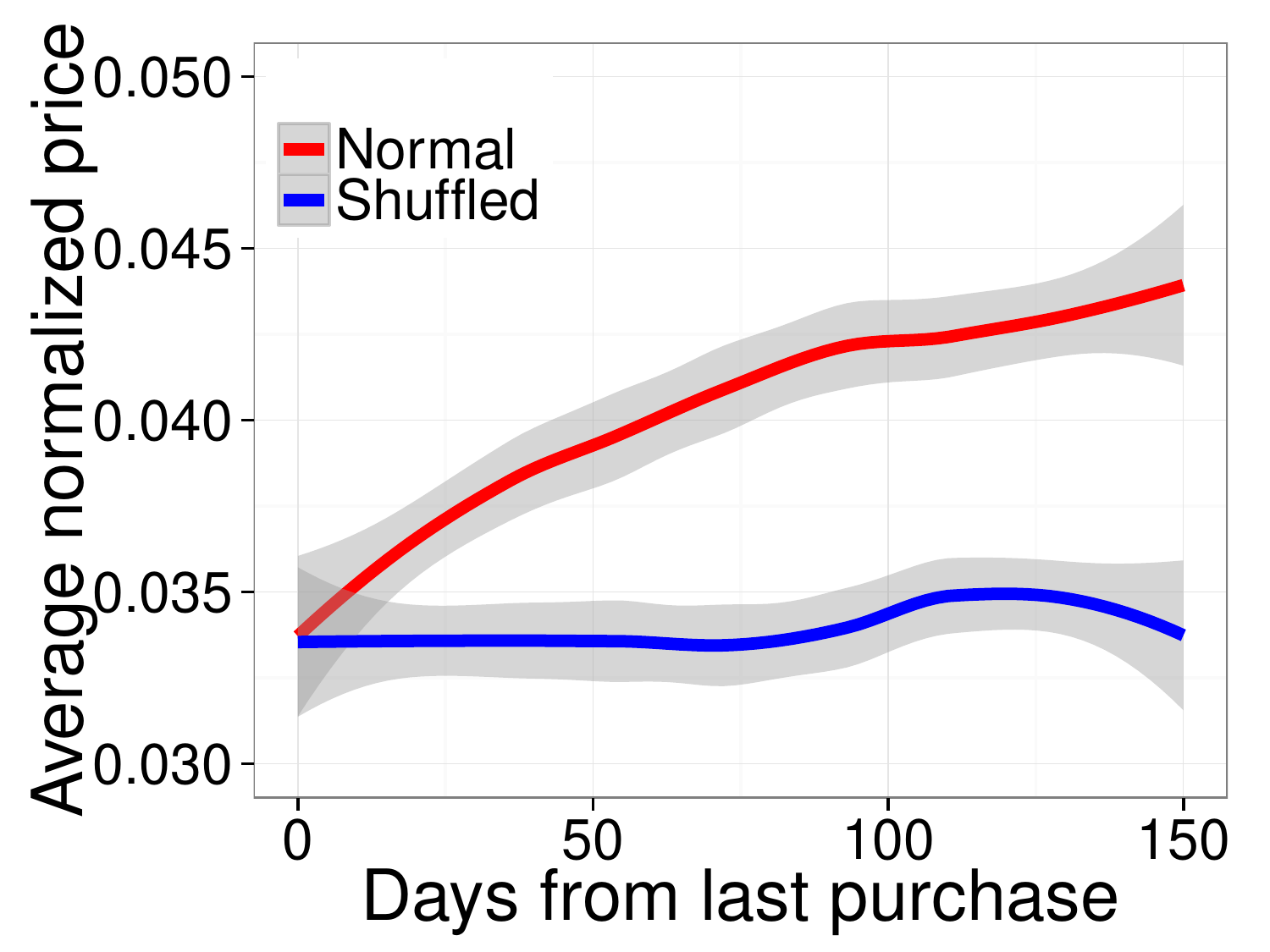}
}

\caption{Relationship between purchase price and time to next purchase with 0.95 confidence interval.}
\vspace{-4mm}
\label{fig:last}
\end{center}
\end{figure*}

An individual's purchasing decisions are not independent, but constrained by their finances. Budgetary constraints introduce temporal dependencies between purchases made by the user: After buying a product, the user has to wait some time to accumulate money to make another purchase. Previous work in economics studied models of budget allocation of households across different types of goods to maximize an utility function~\cite{du08money} and analyzed conditions under which people are willing to break their budget cap~\cite{chiou11spend}. However, we are not aware of any study aimed to support the hypothesis of the time of purchase being partly driven by an underlying cyclic process of budget depletion and replenishment.

To test this hypothesis, we examine the relationship between purchase price and the time period since last purchase. Since different users have different spending power, we consider the normalized change in the price given the number of days from the last purchase. In other words, we compute how users divide their personal spending across different purchases, given the time delay between purchases. We then average the normalized values for all users, and report the change for each time delay. Figure~\ref{fig:last_one} shows that as the time delay gets longer, users spend higher fraction of their budgets, which supports our hypothesis. To test that our analysis does not have any bias in the way the users are grouped we perform a shuffle test, by randomly swapping the prices of products purchased by users. This destroys the correlation between the time delay and product price. We then do the same analysis with the shuffled data and we expect to see a flat line. However, the same increase also exists in the shuffled data, indicating a bias in the methodology. This is due to heterogeneity of the underlying population: we are mixing users with different number of purchases. Users making more purchases have lower normalized prices and also shorter time delays, and they are systematically overrepresented in the left side of the plot, even in the shuffled data.

To partially account for heterogeneity, we grouped users by the number of purchases: e.g., those who made exactly 5 purchases, those with 9-11 purchases, and 28-32 purchases. Even within each group there is variation as the total spending differs significantly across users, which we address by normalizing the product price by the total amount of money spent by the user, as explained above. If our hypothesis is correct, there should be a refractory period after a purchase, with users waiting longer to make a larger purchase. We clearly observe a positive relationship between (normalized) purchase price and the time (in days) since last purchase (Figure~\ref{fig:last}), but not in the shuffled data, which produces a horizontal line. We conclude that the relationship between time delay and purchase price arises due to behavioral factors, such as limited budget.

\subsection{Social Factors}\label{section:diffusion}

\noindent An individual's behavior is often correlated with that of his or her social contacts (or friends). In online shopping, this would result in users purchasing products that are similar to those purchased by their friends. Distinct social mechanisms give rise to this correlation~\cite{Anagnostopoulos08}. First, a friend could influence the user to buy the same product, for example, by highly recommending it. This is the basis for social contagions in general, and ``word-of-mouth'' marketing in particular, although empirical evidence suggests that influence has a limited effect on shopping behavior~\cite{leskovec07dynamics}. Alternatively, users could have bought the same product as their friends purchased before, because people tend to be similar to their friends, and therefore, have similar needs. The tendency of socially connected individuals to be similar is called homophily, and it is a strong organizing principle of social networks. Studies have found that people tend to interact with others who belong to a similar socio-economic class~\cite{Feld1981,mcpherson01bird} or share similar interests~\cite{Kang12,aiello12friendship}. Finally, a user's and friend's behavior may be spuriously correlated because both depend on some other external factor, such as geographic proximity. In reality, all these effects are interconnected~\cite{crandall08feedback,aiello10link} and are difficult to disentangle from observational data~\cite{shalizi11homophily}. For example, homophily often results in selective exposure that may amplify social influence and word-of-mouth effects.

We investigate whether social correlations exist, although we do not resolve the source of the correlation. Specifically, we study whether users who are connected to each other via email interactions tend to purchase similar products in contrast to users who are not connected.
To measure similarity of purchases between two users, we first describe the purchases made by each user with a vector of products, each entry containing the frequency of purchase. This approach results in large and sparse vectors due to the large number of unique products in our data set. To address this challenge, we use vectors of product categories, instead of product names. There are three levels of product categories, and we perform our experiments at all levels.

We compare similarity of category vectors of pairs of users who are directly connected in the email network (104K pairs of users) with the same number of pairs of randomly chosen users (who are not directly connected). We use cosine similarity to measure similarity of two vectors. Using top-level categories to describe user purchases gives average similarity of 0.420 between connected pairs of users, whereas random pairs have similarity of 0.377 on average (+11\% relative change as compared to connected pairs). Using the more detailed level-2 categories to describe purchases gives average similarity of 0.215 for connected versus 0.170 for random pairs of users (+26\% relative increase). Finally using the most detailed, level 3, categories results in average similarity of 0.188 for connected vs 0.145 for random pairs of users (+30\% relative increase). Although the absolute similarity decreases as a more detailed product vector is used, shoppers who communicate by email are always, more similar than random shoppers who are not directly connected.

Gender also plays an important role when measuring purchase similarity between user pairs. To quantify this effect, we calculate the cosine similarity between the vectors of number of purchases from the detailed category (level 3), and take the average of the cosine similarity. Instead of taking the average for all the connected pairs, we separate the pairs based on the gender of the users in the pair: woman-woman, man-man, and woman-man. The woman-woman pairs have the highest average cosine similarity with 0.192, next followed by man-man pairs with average similarity of 0.186. Heterogeneous pairs are the least similar ones, with average cosine similarity of 0.182, still greater than random pairs of users, which have similarity of 0.145, since the users are connected. Woman-man pair having the smallest similarity primarily supports our earlier finding about a sensible difference in the type of goods that attract the interest in the two genders. Previous work also found that receiving a shopping recommendation from a friend will have a greater positive effect on willingness to purchase online among women than among men~\cite{garbarino04gender}. The highest similarity of female-female pairs might be partly determined by that.

\section{Predicting Purchases}\label{section:prediction}

\noindent Predicting the behavior of online shoppers can help e-commerce sites on one hand to improve the shopping experience by the means of personalized recommendations and on the other hand to better meet merchants' needs by delivering targeted advertisements. In a recent study, Grbovic et al. addressed the problem of predicting the next item a user is going to purchase using a variety of features~\cite{grbovic2015commerce}. In this work, we consider the complementary problems of predicting \textit{i)} the \textit{time} of the next purchase and \textit{ii)} the \textit{amount} that will be spent on that purchase.

Predicting the exact time and price of a purchase (e.g., using regression) is a very hard problem, therefore we focused on the simpler classification task of predicting the class of the purchase among a finite number of predefined price or time intervals. We experimented with different classification algorithms and Bayesian Network Classification yielded the highest accuracy. To estimate the conditional probability distributions we used direct estimates of the conditional probability with $\alpha = 0.5$. The classifier was trained on the first six months of purchase data and evaluated on the last two. From each entry we extracted 55 features belonging to variety of categories:
\begin{itemize}[leftmargin=*]
\itemsep0em
\item \textit{Demographics of online shoppers} (4 features): Gender, age, location (zip code), and income (based on zip code)
\item \textit{Purchase price history} (19 features): Price of the last three purchases, price category of the last three purchases, number of purchases, mean price of purchased item, median price of purchased items, total amount of money spent, standard deviation in item prices, number of earlier purchases in each price group (5 groups), price group with the most number of purchases and the count for it, and total number of purchases until that point.
\item \textit{Purchase time history} (13 features): Time of last three purchases, mean time between purchases, median time between purchased, standard deviation in times between purchases, number of earlier purchases in each time group, and time group with the most number of purchases and the count for it.
\item \textit{Purchase history of products} (4): Last three categories of products purchased, most purchased category.
\item \textit{Time or price of the next purchase} (1 feature): We also assume that we know when the next purchase is going to happen. This seems unrealistic at first, but we include this feature because the system is going to make recommendations at a given time, and we assume that the buyer is going to make the decision at that time. For having a symmetrical problem we also consider the price of the next purchase, which would be similar as knowing the budget of the user.
\item \textit{Contacts} (14 features): Mean, median, standard deviation, minimum, maximum, and 10th and 90th percentile of price and time of the purchases of the contacts of the users.
\end{itemize}
For the aggregated features such as average price of item purchased, we used only purchases in the training period and did not consider future information. We compared results of our classifier to three baselines:
\begin{itemize}[leftmargin=*]
\itemsep0em
\item Random prediction.
\item Price or time class of the previous purchase.
\item Most popular price or time class of the target user's earlier purchases.
\end{itemize}
\subsection{Price of the Next Purchase}

We partition prices in five classes using \$6, \$12, \$20, and \$40 as price thresholds to obtain equally-sized partitions. These thresholds represent (a) very cheap products that cost less than \$6 (20.7\% of the data), (b) cheap products between \$6 and \$12 (20.3\%) (c) medium-priced products between \$12 and \$20 (19.3\%), (d) expensive products that cost more than \$20, but less than \$40 (19.9\%), and finally (e) very expensive products worth more than \$40 (19.8\%). Our classifier achieves an accuracy of 31.0\% with a +49.8\% relative improvement over the 20.7\% accuracy of the random classifier (i.e., relative size of the largest class). 

The category of the last purchase and the most frequent purchase category turn out to be quite strong predictors, achieving alone accuracy of 29.3\% and 29.8\%. The supervised approach outperforms them, but with only a +5.8\% and +4.0\% relative improvement respectively. When measuring the predictive power of the features with the $\chi^2$ statistics (Table~\ref{table:feature_rank_price}) we find that the highest predictive power is the most frequent class of earlier purchases, by far. This suggests that users tend to buy mostly items in the same price bracket. The second feature in the ranking is the number of purchases from the very cheap category, followed by median and mean of earlier prices.
In general, all the top 16 most informative features are related to the price of earlier purchases. After those, median time between purchases and time delay before the last purchase are the most predictive features. The relatively high position of the last time delay in the feature rank suggests that the recommender system should consider the time that has passed from the last purchase of the user, and change the suggestions dynamically. In other words, if the user has made a purchase recently, cheaper products should be favored over more expensive products to the user, whereas if a long period of time has passed since the last purchase, more expensive products should be advertised to the user, as they are more likely to be purchased. All of the demographics features have limited predictive power and are ranked last (though the demographics might affect the purchase history), with income being the most important among them.

\begin{table}[t!]
\begin {center}
{
\begin {tabular} {| c | l | r |}
\hline
{\textbf{Rank}} & {\textbf{Feature}} & \textbf{{$\chi^2$} value} \\
\hline
1 & Most used class earlier & 214,996\\
2 & Number of under \$6 purchases  & 115,560\\
3 & Median price of earlier purchases & 106,876\\
4 & Mean price of earlier purchases & 91,409\\
5 & Number of over \$40 purchases & 84,743\\
\hline
\end{tabular}
}
\end{center}
\caption{Top predictive features for prediction of the price of the next item and their {$\chi^2$} value.}
\label{table:feature_rank_price}
\end{table}

\subsection{Time-Between-Interaction of the Next Purchase}
Similarly to purchase price, prediction of purchase time could be leveraged to make a better use of the advertisement space. If the user is likely not to purchase anything for a certain period of time, ads can be momentarily suspended or replaced with ads that are not related to consumer goods.

For creating the categories, we choose thresholds of 1,  5, 14, and 33 days. Very short delays are within a day (22.8\% of our data), short delays between 1 and 5 days (20.9\%), medium delays between 5 and 14 (19.6\%), long delays between 14 and 33 (18.2\%) and the very long delays exceed 33 days (18.5\%). Training a Bayesian Network on all the features yields an accuracy of 31.1\%, a +36.4\% relative improvement over the 22.8\% accuracy of the random prediction baseline. The accuracy of our classifier is also +24.9\% relatively higher than the baseline of predicting as the last purchase delay, which has accuracy of 24.9\%. Finally, the most occurred purchase has an accuracy of 22.2\%, which is outperformed by our classifier by +40.1\% relatively.

Ranking features by their $\chi^2$ (Table~\ref{table:feature_rank_time}), we find that the most informative feature is the number of earlier purchases that the user has made so far, followed by median time delay, previous purchase delay, time since the first purchase, and the class of the previous purchase delay.

\begin{table}[t!]
\begin {center}
{
\begin {tabular} {| c | l | r |}
\hline
{\textbf{Rank}} & {\textbf{Feature}} & \textbf{{$\chi^2$} value} \\
\hline
1 & Number of earlier purchases & 48,719\\
2 & Median time between purchases & 35,558\\
3 & Time since the first purchase & 30,741\\
4 & Previous time delay & 30,692\\
5 & Class of previous time delay & 22,710\\
\hline
\end{tabular}
}
\end{center}
\caption{Top predictive features for prediction of time of next purchase and their {$\chi^2$} value.}
\label{table:feature_rank_time}
\end{table}

To summarize, we trained two classifiers for predicting the price and the time of the next purchase. Our algorithm outperformed the baselines in both prediction tasks, by a higher margin in case of predicting the time. Table~\ref{table:prediction_results} summarizes all of our results showing a relative improvement of at least 49.8\% for predicting the price of the next item purchased and 36.4\% for predicting the time of the next purchase over the majority vote baseline. Interestingly, user demographics were not particularly helpful for making any prediction, and the observed correlations in earlier sections of the paper are masked by other features such as the history of prior purchases.

\begin{table*}[t!]
\begin {center}
{
\begin {tabular} {| l | c | c  | c | c | c | c | c | c |}
\hline
{Prediction} &  \specialcell{Majority vote \\ (random classifier)} & Last used & {Most used} & {Our classifier} &  \specialcell{Absolute \\ improvement} & \specialcell{Relative \\ improvement} & AUC & RMSE\\
\hline
Item price        & 20.7\% & 29.3\% & 29.8\% & 31.0\% & 10.3\% & 49.8\% & 0.611 & 0.4641 \\
Purchase time & 22.8\% & 24.9\% & 22.2\% & 31.1\% & 8.3\% & 36.4\% & 0.634 & 0.4272\\
\hline
\end{tabular}
}
\end{center}
\caption{Summary of the prediction results. Accuracy: percentage of correctly classified samples. Majority vote: always predicting the largest group, or predicting randomly. Most used: the group the user had the most in earlier purchases. AUC: Weighted average of Area Under the Curve for classes. RMSE: Root Mean Square Error. The improvements are reported over the majority vote baseline.}
\vspace{-2mm}
\label{table:prediction_results}
\end{table*} 

\section{Related Work}\label{section:related}

\noindent Most of previous research on shopping behavior and characterization of shoppers has been conducted through interviews and questionnaires administered to groups of volunteers composed by at most few hundred members.

\textbf{Offline shopping} in physical stores has been studied in terms of the role of demographic factors on the attitude towards shopping. The customer's gender predicts to some extent the type of purchased goods, with men shopping more for groceries and electronics, while women more for clothing~\cite{dholakia99going,JOCA:JOCA113}. Gender is also a discriminant factor with respect to the attitude towards financial practices, financial stress, and credit, and it can be a quite good predictor of spending~\cite{JOCA:JOCA113}. Many shoppers express the need of alternating the experience of online and offline shopping~\cite{wolfinbarger01shopping,tabatabaei09online}, and it has been found that there is an engagement spiral between online and offline shopping: searching for products online positively affects the frequency of shopping trips, which in its turn positively influences buying online~\cite{farag07shopping}.

\textbf{Online shopping} has been investigated since the early stages of the Web. Many studies tried to draw the profile of the typical online shopper. Online shoppers are younger, wealthier, more educated than the average Internet user. In addition, they tend to be computer literate and to live in the urban areas~\cite{zaman02internet,swinyard03people,swinyard11activities,farag07shopping}. Their trust of e-commerce sites and their understanding of the security risks associated with online payments positively correlate with their household income and education level~\cite{horrigan2008online,hui07factors}, and it tends to be stronger in males~\cite{garbarino04gender}. The perception of risk of online transactions influences shoppers to purchase small, cheap items rather than expensive objects~\cite{bhatnagar00risk}. Customers of online stores tend to value the convenience of online shopping in terms of ease of use, usefulness, enjoyment, and saving of time and effort~\cite{perea04drives}. Their shopping experience is deeply influenced by their personal traits (e.g., previous online shopping experiences, trust in online shopping) as well as other exogenous factors such as situational factors or product characteristics~\cite{perea04drives}.

\textbf{Demographic factors} can influence the shopping behavior and the perception of the shopping experience online. Men value the practical advantages of online shopping more and consider a detailed product description and fair pricing significantly more important than women do. In contrast, some surveys have found that women, despite the ease of use of e-commerce sites, dislike more than men the lack of a physical experience of the shop and value more the visibility of wide selections of items rather than accurate product specifications~\cite{vanslyke02gender,vanslyke05gender,ulbrich11gender,hui07factors}. Unlike gender, the effect of age on the purchase behavior seems to be minor, with older people searching less for online items to buy but not exhibiting lower purchase frequency~\cite{sorce05attitude}. With extensive evidence from a large-scale dataset we find that age greatly impacts the amount of money spent online and the number of items purchased.

\textbf{The role of the social network} is also a crucial factor that steers customer behavior during online shopping. Often, social media is used to communicate purchase intents, which can be automatically detected with text analysis~\cite{gupta14identifying}. Also, social ties allow for the propagation of information about effective shopping practices, such as finding the most convenient store to buy from~\cite{guo11role} or recommending what to buy next~\cite{leskovec07dynamics}. Survey-based studies have found that shopping recommendations can increase the willingness of buying among women rather than men~\cite{garbarino04gender}.

\textbf{Factors leading to purchases} in offline stores have been extensively investigated as they have direct consequences on the revenue potential of retailers and advertisers. Survey-based studies attempted to isolate the factors that lead a customer to buy an item or, in other words, to understand what the strongest predictors of a purchase are. Although the mere amount of online activity of a customer can predict to some extent the occurrence of a future purchase~\cite{bellman99predictors}, multifaceted predictive models have been proposed in the past. Features related to the phase of information gathering (access to search features, prior trust of the website) and to the purchase potential (monetary resources, product value) can often predict whether a given item will be purchased or not~\cite{hansen04predicting,pavlou06understanding}.

\textbf{Prediction of purchases in online shopping} is a task that has been addressed through data-driven studies, mostly on click and activity logs. User purchase history is extensively used by e-commerce websites to recommend relevant products to their users~\cite{linden03amazon}. Features derived from user events collected by publishers and shopping sites are often used in predicting the user's propensity to click or purchase~\cite{djuric14hidden}. For example, clickstream data have been used to predict the next purchased item~\cite{vandenpoel05predicting,senecal05consumers}; click models predict online buying by linking the purchase decision to what users are exposed to while on the site and what actions they perform while on the site~\cite{montgomery02predicting,sismeiro04modeling}. Besides user click and purchase events, one can leverage product reviews and ratings to find relationships between different products~\cite{mcauley2015reviews}. Email is also a valuable source of information to analyze and predict user shopping behavior~\cite{grbovic2015commerce}. Click and browsing features represent only a weak proxy of user's purchase intent, while email purchase receipts convey a much stronger intent signal that can enable advertisers to reach their audience. The value of commercial e-mail data has been recently explored for the task of clustering commercial domains~\cite{grbovic14generating}. Signals to predict purchases can be strengthen by demographic features~\cite{kim03combination}. Also, the information extracted from the customers' profiles in social media, in combination with the information of their social circles, can help with predicting the category of product that will be purchased next~\cite{zhang13predicting}.

\section{Conclusion}\label{section:conclusion}

\noindent

Studying the online consumer behavior as recorded by email traces allows to overcome the limitations of previous studies that focused either on small-scale surveys or on purchases' logs from individual vendors. In this work, we provide the first very large scale analysis of user shopping profiles across several vendors and over a long time span. We measured the effect of age and gender, finding that the spending ability goes up with age till the age of 30, stabilizes in the early 60s, and then starts dropping afterwards. Regarding the gender, a female email user is more likely to be an online shopper than an average male email user. On the other hand, men make more purchases, buy more expensive products on average, and spend more money. Younger users tend to buy more phone accessories compared to older users, whereas older users buy TV shows and vitamins \& supplements more frequently. Using the user location, we show clear correlation between income and the number of purchases users make, average price of products purchased, and total money spent. Moreover, we study the cyclic behavior of users, finding weekly patters where purchases are more likely to occur early in the week and much less frequently in the weekends. Also, most of the purchases happen during the work hours, morning till early afternoon.

We complement the purchase activity with the network of email communication between users. Using the network, we test if users that communicate with each other have more similar purchases compared to a random set of users, and we find indeed that is the case. We also consider gender of the users and find that woman-woman pairs are more similar than man-man pairs that are also more similar to each other than the woman-man pairs. Finally, we use our findings to build a classifier to predict the price and the time of the next purchase. Our classifier outperforms the baselines, especially for the prediction of the time of the next purchase. This classifier can be used to make better recommendations to the users.

Our study comes with a few limitations, as well. First, we can only capture purchases for which a confirmation email has been delivered. We believe this is the case for most of online purchases nowadays. Second, if users use different email addresses for their purchases, we would not have their full purchase history. 
Similarly, people can share a purchasing account to enjoy some benefits (e.g., an Amazon Prime account between multiple people), but that occurs rarely, as suggested by the fact that less than 0.01\% of the users have goods shipped to more than one zip-code. Third, the social network that we considered, albeit big, is not complete. However, the network is large enough to observe statistically significant results. Lastly, we considered the items that were purchased together as separate purchases; it would be interesting to see which items are usually bought together in the same transaction.

\balance

\end{document}